\documentclass{mn2e}
\usepackage[utf8x]{inputenc}
\usepackage{float}
\usepackage{graphicx}
\usepackage{amssymb}
\usepackage{amsmath}
\usepackage[usenames,dvipsnames]{xcolor}
\usepackage{astrojournals}
\usepackage{multicol}
\usepackage{multirow}
\usepackage{hyperref}
\usepackage{enumerate}
\usepackage{natbib}
\newcommand{\mvir}{M_\mathrm{v}}
\newcommand{\rvir}{R_\mathrm{v}}
\newcommand{\vvir}{V_\mathrm{v}}
\newcommand{\vmax}{V_\mathrm{max}}
\newcommand{\vpeak}{V_\mathrm{peak}}
\newcommand{\msub}{M}
\newcommand{\msun}{M_\odot}
\newcommand{\mstar}{M_{\star}}
\newcommand{\mpeak}{M_\mathrm{peak}}
\newcommand{\apeak}{a_\mathrm{peak}}

\newcommand{\mpc}{\mathrm{Mpc}}
\newcommand{\kpc}{\mathrm{kpc}}

\newcommand{\kms}{{\rm km} \, {\rm s}^{-1}}
\newcommand{\lcdm}{$\Lambda$CDM}

\title{ELVIS:  Exploring the Local Volume in Simulations}
\author[S. Garrison-Kimmel et al.]{Shea Garrison-Kimmel$^1$\thanks{$\!$sgarriso@uci.edu},
  Michael Boylan-Kolchin$^{2,1}$,
  James S. Bullock$^1$,
  Kyle Lee$^3$\\
  \noindent$\!\!$ $^1$Center for Cosmology, Department of Physics and Astronomy,
  University of California, Irvine, CA 92697, USA \\
  \noindent$\!\!$ $^2$Department of Astronomy and Joint Space-Science Institute,
  University of Maryland, College Park, MD 20742-2421, USA \\
  \noindent$\!\!$ $^3$Chapman University, Orange, CA 92866, USA}
\begin{document}

 \pagerange{\pageref{firstpage}--\pageref{lastpage}} 
 \pubyear{2013}

\maketitle

\label{firstpage}
\begin{abstract}
  We introduce a set of high-resolution dissipationless simulations that model
  the Local Group (LG) in a cosmological context: Exploring the Local Volume in
  Simulations (ELVIS).  The suite contains 48 Galaxy-size haloes, each within
  high-resolution volumes that span $2-5$ Mpc in size, and each resolving
  thousands of systems with masses below the atomic cooling limit.  Half of the
  ELVIS galaxy haloes are in paired configurations similar to the Milky Way (MW)
  and M31; the other half are isolated, mass-matched analogs.  We find no
  difference in the abundance or kinematics of substructure within the virial
  radii of isolated versus paired hosts. On Mpc scales, however, LG-like pairs
  average almost twice as many companions and the velocity field is
  kinematically hotter and more complex.  We present a refined abundance
  matching relation between stellar mass and halo mass that reproduces the
  observed satellite stellar mass functions of the MW and M31 down to the regime
  where incompleteness is an issue, $\mstar \sim 5\times 10^5 \, \msun$.  Within
  a larger region spanning approximately $3\, \mpc$, the same relation predicts
  that there should be $\sim 1000$ galaxies with $\mstar > 10^{3}\,\msun$
  awaiting discovery.  We show that up to $50\%$ of haloes within $1$ Mpc of the
  MW or M31 could be systems that have previously been within the virial radius
  of either giant.  By associating never-accreted haloes with gas-rich dwarfs, we
  show that there are plausibly $50$ undiscovered dwarf galaxies with HI masses
  $> 10^5 \msun$ within the Local Volume. The radial velocity distribution of
  these predicted gas-rich dwarfs can be used to inform follow-up searches based
  on ultra-compact high-velocity clouds found in the ALFALFA survey.
\end{abstract}


\section{Introduction}
\label{sec:intro}

The Local Group (LG) provides an important test bed for theories of galaxy
formation, both in its connection to small-scale probes of the consensus dark
energy plus cold dark matter (\lcdm) cosmological model \citep{klypin1999,
  moore1999, strigari2008, walker2011, Boylan-Kolchin2011, Zolotov2012,
  Garrison-Kimmel2013, Arraki2013} and as a potential hunting ground for the
descendants of reionization and first light
\citep{bullock2000,ricotti2005,Madau2008}.  The focus on these issues is
well motivated: given inevitable completeness limitations, nearby galaxies offer
our best avenue for characterizing the faint end of the global luminosity
function and for studying resolved stellar populations as beacons from an
earlier age \citep[see, e.g.][]{Makarov2011, Weisz2011, McConnachie2012,
  Karachenstsev2013, Tully2013}.

Numerical simulations have emerged as the most useful tool for making
predictions about non-linear structures in \lcdm. While simulations of
cosmologically large volumes enable statistical comparisons with a variety of
observations \citep[e.g.][]{Davis1985, Gross1998, Springel2005b,
  Boylan-Kolchin2009, Klypin2011}, cosmological zoom-in simulations are the de
facto standard for the most detailed comparisons of individual objects.  The
zoom-in technique \citep{Katz1993,Onorbe2013} focuses computational power on a
small, high resolution region nested within a lower resolution,
cosmological-size volume, thereby retaining the large-scale, low frequency
cosmological modes important for convergence but also allowing for the high
resolutions required to obtain a wide dynamic range.

Very high resolution zoom-in simulations of Milky Way (MW) mass haloes have 
been useful for making and testing predictions of the $\Lambda$CDM theory
\citep[e.g.][]{Diemand2008,Kuhlen2008,Aquarius}, often through comparisons to
dwarf satellite galaxies of the LG \citep{koposov2009,Strigari2010,Boylan-Kolchin2012}.  
However, in order to achieve the highest resolution possible, these simulations have 
concentrated on fairly isolated systems.\footnote{As noted in \citet{Teyssier2012}, the 
  Via Lactea II halo does indeed have a massive ($\mvir=6.5\times10^{11}\,\msun$)
  companion at a distance comparable to M31. However, this companion halo and
  field galaxies nearby are not free of contamination from low-mass
  particles. The contamination reaches $50\%$ by mass, which has potentially
  important effects on halo properties.} In reality, the Milky Way is not
isolated, but has a nearby companion of comparable luminosity, the Andromeda
galaxy (M31).  The existence of M31 at a distance of approximately 800 kpc from
the MW implies that isolated zoom simulations cannot be used to faithfully make
predictions for the Local Volume\footnote{A term we use roughly to correspond to 
a $\sim 2$ Mpc sphere from the LG barycenter.} beyond $\sim 400$ kpc of either system.
Furthermore, several simulations that have explored the role of LG-like
environments in shaping galaxy properties have found evidence that the local
configuration may even bias galaxy properties within each giant's virial radius
compared to isolated counterparts \citep{Gottloeber2010, Libeskind2010,
  Forero-Romero2011, Few2012}.

Motivated by these concerns, here we introduce a set of dissipationless
simulations designed to confront the Local Volume in a cosmological context.  
We call this project Exploring the Local Volume in Simulations (ELVIS).  The 
simulation suite consists of 12 zoom-in regions of LG analogue halo pairs and 
24 isolated haloes that are mass-matched to create a control sample for those pairs.  
Below, we describe the selection of ELVIS pairs, their simulation details, and 
properties of the host haloes (\S~\ref{sec:sims}). We investigate the environments 
that surround them in comparison to those of the control hosts as well as the 
dynamical histories of bound haloes around the ELVIS giants by characterizing the 
fraction of  `backsplash' haloes -- systems that at one point had been within the virial
radius of a giant -- as a function of distance (\S~\ref{sec:LGvIso}). Finally,
we compare number counts and kinematic properties of the subhaloes found in
paired and isolated samples, and we use abundance matching (AM) to make predictions
for the stellar and HI mass functions within the Local Volume
(\S~\ref{sec:predictions}).

With the publication of this paper, we publicly release all of the data in the
ELVIS suite, including full merger trees, $z = 0$ halo catalogs, and particle
information.\footnote{Present-day ($z = 0$) halo catalogs and the main branches
  of the merger trees are available for public download 
  (\url{http://localgroup.ps.uci.edu/elvis}), while access to 
  the full merger trees and particle information will be arranged via email contact 
  with the authors.}

\section{The ELVIS Suite}		
\label{sec:sims}

The ELVIS simulations were run using \texttt{GADGET-3} and \texttt{GADGET-2}
\citep{Springel2005}, both tree-based $N$-body codes.  For the underlying
cosmological model, we have adopted \lcdm\ parameters set by the WMAP-7 results
\citep{Larson2011}: $\sigma_8 = 0.801$, $\Omega_m = 0.266$, $\Omega_\Lambda =
0.734$, $n_s = 0.963$, and $h = 0.71$.  Throughout this work, we use the term
virial mass $\mvir$ to refer to the mass within a sphere of radius $\rvir$
that corresponds to an over density of $97$ relative to the critical density of
the Universe \citep{Bryan1998}.  All simulations were initialized at redshift
$z=125$ unless otherwise specified.

\subsection{Halo Selection}
\label{subsec:selection}
We select LG-like pairs from 50 medium-resolution cosmological simulations,
each a cubic volume $70.4$ Mpc on a side with particle mass $m_{\rm p} =
9.7\times10^7\msun$ and Plummer-equivalent force softening length 1.4~kpc
(comoving).  From these cosmological volumes, we selected 12 halo pairs for
resimulation using the criteria described below.  For each of the 24 haloes
included in the ELVIS pairs, we also chose an isolated analogue of identical
virial mass ($\mvir$) that is separated by at least 2.8 Mpc from all haloes more
massive than $\mvir/2$.  The isolated set serves as a control sample for
comparison.

Our approach to selecting LG regions differs from that of the well-known
Constrained Local Universe Simulations (CLUES) project \citep{Gottloeber2010}, 
which relies on the `constrained realization' technique to match the observed 
density and velocity fields on a $\sim 10\,\mpc$ scales around the LG.  The 
advantage to our approach is that it guarantees a good LG analogue in each re-simulation.  
The downside is that the larger scale density field will usually not be identical 
to that of the LG.  The two initialization methods therefore have different,
complementary strengths.

In selecting pairs, we targeted haloes with phase-space characteristics similar
to the MW/M31 system, with cuts similar to those of \citet{Forero-Romero2011},
based on values of the virial mass of each host ($M_\mathrm{v,1}$ and
$M_\mathrm{v,2}$, where $M_\mathrm{v,2} \ge M_\mathrm{v,1}$), the
distance between host centers $\Delta R$, the pair approach velocity, and
local environment:

\begin{itemize}
\item \textit{Mass of each host:} 
		 $10^{12} \leq \mvir \leq 3\times 10^{12}\msun$ \\
                \indent \citep{Tollerud2012, MBK2013, Fardal2013,Piffl2013} \\
\item \textit{Total mass:} 
		$2 \times 10^{12} \leq M_\mathrm{v,1} + M_\mathrm{v,2} \leq 5\times 10^{12}\msun$ \\
		 \indent  \citep{LW08,Marel2012} \\
\item \textit{Separation:}
		 $0.6 \leq \Delta R \leq 1\,\mathrm{Mpc}$ \\
		\indent  \citep[][and references therein]{McConnachie2005} \\
\item \textit{Radial velocity:}
		$V_\mathrm{rad} \leq 0$ km/s \\ 
		\indent \citep{Marel2012} \\
      \item \textit{Isolation:} No haloes with $\mvir \geq M_\mathrm{v,1}$ within
        2.8 Mpc of either centre and no haloes with $\mvir \geq 7\times 10^{13}
        \msun$ within $7$ Mpc of either centre
        \citep{Tikhonov2009,Karachentsev2004}.
\end{itemize}

We identified $146$ halo pairs that met these criteria within the 50
simulations we ran (an equivalent volume of $1.76 \times 10^{7}$ Mpc$^3$) and
selected 12 pairs for resimulation.  We intentionally chose several pairs
that consisted of hosts with massive ($V_{\rm max} > 75$ $\kms$) subhaloes in
order to ensure that we had a fair number of systems with realistic analogs to
the Large Magellanic Cloud (LMC) and M33; had we selected pairs at random, it 
would have been unlikely to obtain such massive subhaloes \citep{Boylan-Kolchin2010}.  
We further made an effort to include two pairs that had very low relative tangential 
velocities $<15~\kms$ in order to mimic the low relative tangential speed of the MW/M31 
pair \citep{Marel2012}.  For the isolated control sample, we imposed no selection
choices other than in matching virial masses and demanding that there are no
haloes with $M > M_{\rm v}/2$ within 2.8 Mpc.  Most of the matches in mass are
good to within $5 \%$, though some differ by up to 10\%.  Though we attempted
to match their masses at the percent level in the low-resolution simulations
used to identify objects for resimulation, differences of this order are
expected when using the zoom-in technique \citep{Onorbe2013}.
 
For record-keeping purposes, each LG-analogue pair is named after a famous duo, as
summarized in Table~\ref{tab:pairprops}.  The individual haloes that make up the
pairs are referenced by the same names in Table~\ref{tab:solopairprops}.  The
isolated analogs are identified by the same name prefixed by $i$ in
Table~\ref{tab:isoprops}.  We discuss the information presented in these tables
in Section~\ref{subsec:properties}.  The first pair listed in Table~\ref{tab:pairprops},
Zeus \& Hera, is singled-out in several figures below as a good analogue to the
M31/MW system in terms of observed galaxy counts in the Local Volume
region.  The halo Hera is identified with the MW in this pairing.

\subsection{Zoom Simulations}
\label{subsec:numerics}

In creating the zoom-in initial conditions for the ELVIS haloes, we broadly
followed the methods outlined in \citet{Onorbe2013}, who give prescriptions for
selecting regions that will be free from low-resolution particle contamination
in the final run.  For the pairs, we identified Lagrangian volumes for all
particles within $4\,\rvir$ of either host in the final timestep; for the
isolated analogs, we use particles within $5\,\rvir$ in all but one case
(specified below).  We relied on the public\footnote{The link is \url{http://www.phys.ethz.ch/~hahn/MUSIC/}} 
code \texttt{MUSIC} \citep{Hahn2011}
to create initial conditions associated with these Lagrangian volumes at high
resolution.  The mass resolution in the zoom regions of our production runs is
$m_p = 1.9\times 10^5 \msun$, corresponding to an effective resolution of
$4096^3$ in the box.  The Plummer-equivalent force softening, $\epsilon$, in
these runs was held constant in comoving units until $z=9$, at which
point it was held fixed at 141~pc (physical) for the remainder of each simulation.

The high resolution regions are surrounded by stepped levels of progressively
lower force resolution and higher mass particles, with the majority of the
parent boxes ($70.4$ Mpc cubes) filled with an effective resolution of $128^3$
($m_p = 6.2\times 10^9\msun$) and each successive step increasing the effective
resolution by a factor of $2$ (decreasing the particle mass by a factor of $8$).
As in the high resolution regions, $\epsilon$ remains constant in comoving units
until $z = 9$, then becomes fixed in physical units.  These force softenings,
however, are significantly larger than in the high resolution regions: at $z =
0$ in the main runs, the two highest particle masses utilize $\epsilon =
56$~kpc, the two intermediate regions use $\epsilon = 4.2$~kpc, and $\epsilon =
704$~pc for the particles immediately surrounding the high resolution volume.

Self-bound dark matter clumps are identified with the six-dimensional halo
finder \texttt{Rockstar} \citep{Behroozi2013} and followed through cosmic time
with \texttt{Consistent-Trees} \citep{Behroozi2013b}.  Both of these codes are
publicly available.\footnote{The links are
  \url{http://code.google.com/p/rockstar/} and
  \url{http://code.google.com/p/consistent-trees/}.}  Subhalo masses
($\msub$) are calculated by \texttt{Rockstar} and correspond to the bound
mass of the system.  Maximum circular velocities ($\vmax$) correspond to the
peak of the circular velocity curve, $V_c(r) = \sqrt{G M(r)/r}$, at a given
redshift.  We also checked results at the final timestep ($z=0$) against the
public\footnote{The link is \url{http://popia.ft.uam.es/AHF/}}, spherical
overdensity-based \texttt{Amiga Halo Finder} \citep{Knollmann2009} and found
that the results did not differ significantly and were identical within the
statistical variation of our sample of haloes.\footnote{Though our results
  presented here and made publicly available upon publication rely on
  \texttt{Rockstar}, we are happy to supply associated \texttt{Amiga Halo
    Finder} catalogs upon request.}

\begin{figure*}		
  \centering
  \includegraphics[width=0.495\textwidth]{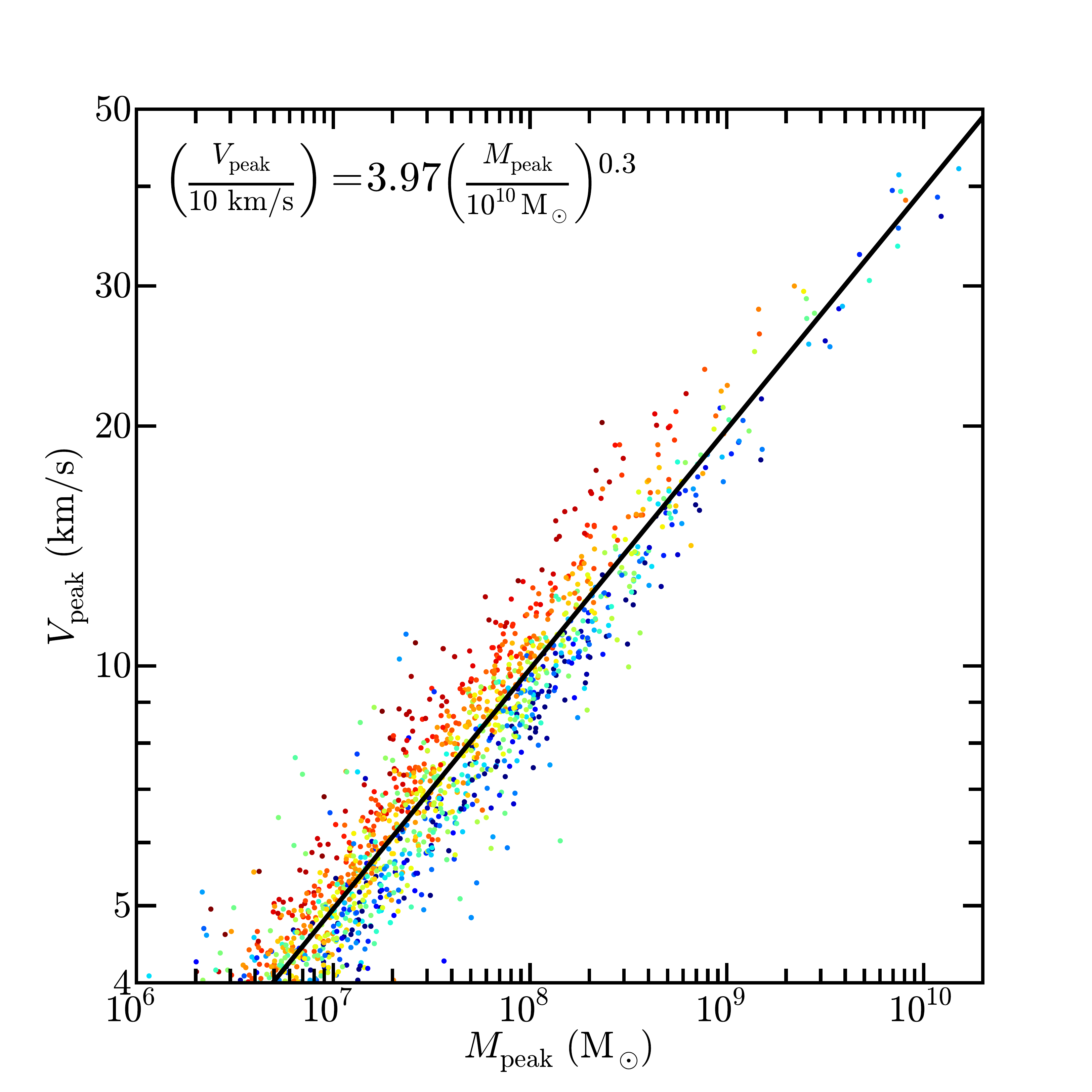}
  \centering
  \includegraphics[width=0.495\textwidth]{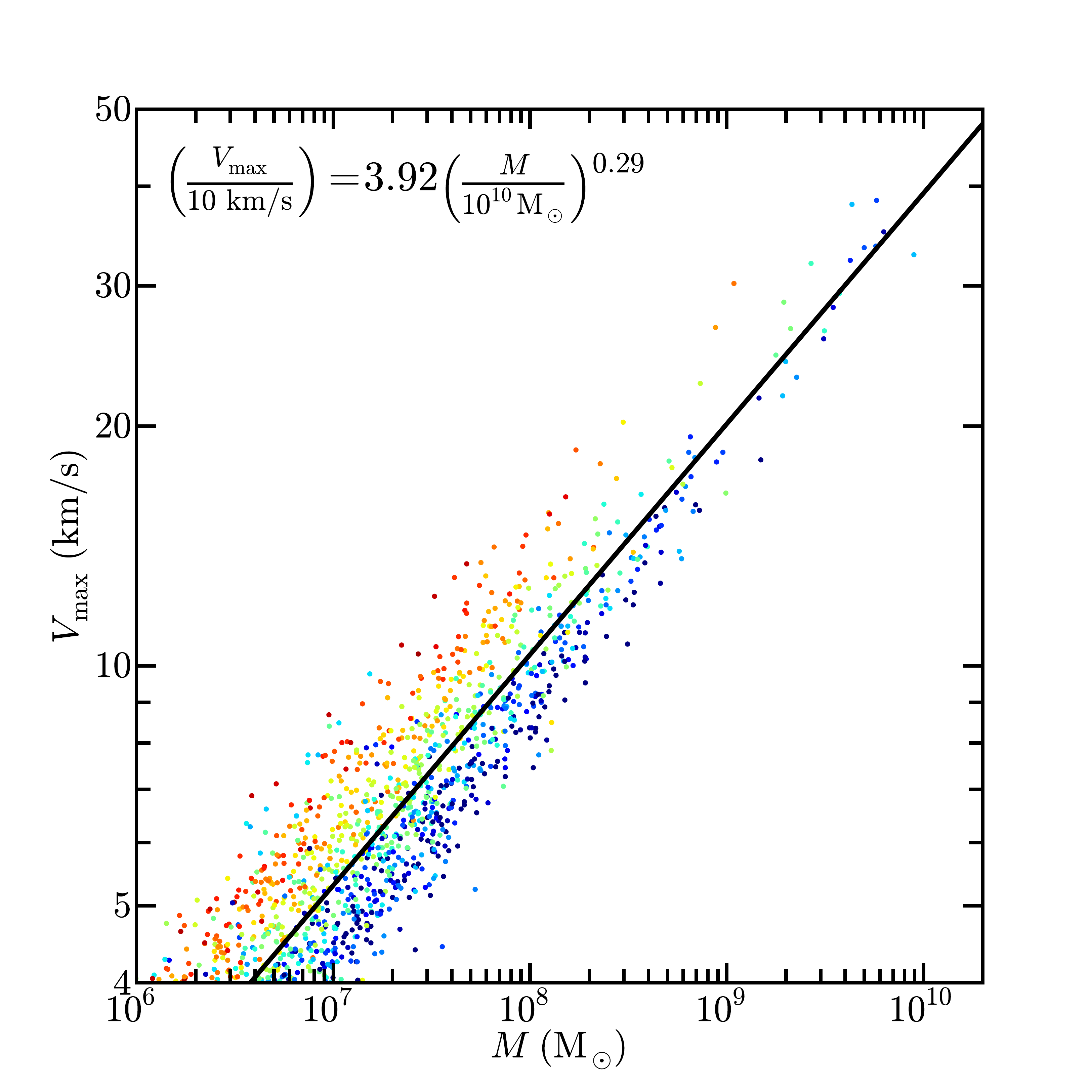}
  \centering
  \includegraphics[width=0.98\textwidth]{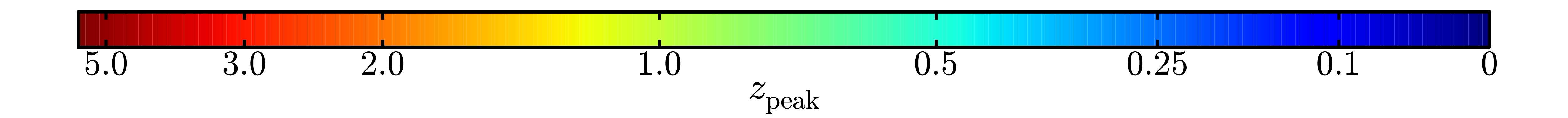}
  \caption{The relation between peak circular velocity and mass at $a_{\rm
      peak}$ (left) and at $z = 0$ (right) The indicated fit includes all
    resolved haloes within $400$~kpc of iKauket in the HiRes run (though the
    results do not differ at the fiducial resolution). Each subhalo is coloured
    by the redshift at which it reached its peak mass ($a_{\rm peak}$); this
    quantity is well-correlated with the scatter about the fits and, as is
    evident from the right panel, the amount of tidal stripping each subhalo has
    undergone.}
  \label{fig:Mpeakrelation}
\end{figure*}

Three of the most useful quantities that can be determined for haloes in our
simulations are $\mpeak$ (the maximum mass of a dark matter structure over its
history), $a_{\rm peak}$ (the latest scale factor at which $\mpeak$ occurs), and
$\vpeak$ (the maximum circular velocity at $a_{\rm peak}$). To define these
quantities, one must adopt an unambiguous definition of the main branch of each
halo's merger tree.  We assign the main progenitor at each timestep as the
branch of the tree with the most total mass up to and including that timestep,
i.e., the sum of $\mvir$ for all haloes over all preceding timesteps in that
branch.  This definition weights both the formation time and the virial mass of
haloes in a given branch. The final step in our pipeline identifies the main
branch of each merger tree and extracts $\mpeak$, $\vpeak$, and $a_{\rm peak}$
for each halo with $z=0$ quantities $\msub$ (or $\mvir$ for hosts) and
$\vmax$.

We simulated three of the isolated analogs (iScylla, iKauket, and iHall) at
higher resolution, with $m_p = 2.35 \times 10^4 \msun$ ($8192^3$ effective
particle number) and $\epsilon=70.4\,{\rm pc}$; we refer to these runs as the
HiRes simulations. The HiRes version of iKauket was originally simulated in the
context of previous work \citep{Onorbe2013} and was initialized at a different
redshift ($z = 250$) than the rest of our runs. It also used a Lagrangian volume
chosen from all particles within $2 \, \rvir$ (rather than our fiducial $5 \,
\rvir$ for the other isolated systems).  The standard resolution version of
iKauket also started at $z = 250$. As \citet{Onorbe2013} showed, any variations
in halo properties at low redshifts introduced by such a change in initial redshift are
comparable to expected variations upon resimulation due to numerical
``minichaos'' \citep{Miller1964}, which should be unimportant for our purposes.

Our HiRes simulations are comparable in mass and force resolution to the
Aquarius level 2 simulations \citep{Aquarius} and to Via Lactea I \citep{VL1}, 
though two of them (iScylla and iHall) have uncontaminated high-resolution volumes --
uncontaminated spheres of radius $\sim 1.5$ Mpc around each host -- that extend
much farther from the halo centers than any previous runs of this kind.  The
HiRes simulations will facilitate several inquiries that are not possible with
our fiducial runs, but for the purposes of this paper, they have allowed us to
self-consistently identify the completeness limit for subhaloes in our main
ELVIS suite.  We find that we are complete to $\msub > 2 \times 10^7\,\msun$,
$\vmax >8\,\kms$, $\mpeak > 6 \times 10^7 \, \msun$ and $\vpeak > 12 \, \kms$.
The numerical convergence of our results in $\vmax$ and $\mpeak$ is demonstrated
explicitly for iKauket in Appendix~A.

In the bulk of this paper, we will enumerate haloes and subhaloes based on
$\mpeak$.  One could equivalently present results in terms of
$\msub$, $\vmax$, or $\vpeak$ ($\vmax$ functions are presented Appendix~B).
Figure~\ref{fig:Mpeakrelation} demonstrates the relationship between $\mpeak$
and $\vpeak$ (left panel) and $\msub$ and $\vmax$ (right panel) for haloes
within 400~kpc of the HiRes version of iKauket (the results are
indistinguishable for the fiducial resolution runs for $\vmax >8\,\kms$ and
$\mpeak > 6 \times 10^7~\msun$.)  The best-fitting $\mpeak - \vpeak$ and $\msub
- \vmax$ relations are given by the formulas in the figures themselves.

What is the origin of the scatter in the $V-M$ relations?  The points in
Figure~\ref{fig:Mpeakrelation} are coloured by $z_{\rm peak} = a_{\rm peak}^{-1}
- 1$.  We see that this variable is strongly correlated with the scatter in $V$
at fixed $M$, such that earlier-forming haloes have higher values of $\vpeak$ and
$\vmax$. The correlation between $\apeak$ and $\vpeak$ is related to the
redshift dependence of the virial overdensity.  At early times, haloes at
fixed mass have a higher $\vmax$.  The red points effectively sample
a population of haloes at $z > 3$, whereas the blue points correspond
to haloes in the field at $z \lesssim 0.1$.  The correlation between $\apeak$ 
and $\vmax$ is a combination of the $\apeak-\vpeak$ correlation and the 
effects of orbital evolution on subhalo density structure \citep[for discussions 
on these expected trends see, e.g.,][]{zb03,Kazantzidis04,dkm07}.

\renewcommand{\arraystretch}{0.75}
\newcommand{\minitab}[2][l]{\begin{tabular}{#1}#2\end{tabular}}
\begin{table*}	
\begin{center}
\setlength{\tabcolsep}{3pt}
\begin{tabular}{|l|c|c|c|c|c|c|c|c|c}
\hline\textbf{Pair Name} & $\Delta R$ & \textbf{V}$_\mathrm{rad}$  &
\textbf{V}$_\mathrm{tan}$ & Total Mass $^a$ & Mass Ratio $^b$ &
$\mathbf{\mathcal{V}}_\mathrm{res}$ $^c$ &
\textbf{N}$_\mathrm{haloes}$ $^d$ & 
\textbf{N}$_\mathrm{p}$ $^e$   & 
[\textbf{V}$_\mathrm{3}$, \textbf{D}$_\mathrm{\ell}$, \textbf{D}$_\mathrm{s}$] $^f$ \smallskip \\ 
 & (kpc) & (km/s) & (km/s) & ($10^{12}M_\odot$) & 
 & ($\mathrm{Mpc^3}$) & 
 ($< \mathcal{V}_\mathrm{res}$) & 
 ($< \mathcal{V}_\mathrm{res}$) &
 (km/s, Mpc, Mpc) \\ \hline\hline 
Zeus \& Hera & 595 & -158.4 & 173.6 & 3.98 & 2.05 & 39.7 & 3,956 & 44M & [73, 0.73, 1.3] \\ \hline
Scylla \& Charybdis & 705 & -21.1 & 132.4 & 3.97 & 1.45 & 38.1 & 4,381 & 47M & [105, 0.50, 1.09] \\ \hline
Romulus \& Remus & 935 & -20.4 & 13.2 & 3.15 & 1.53 & 34.6 & 2,522 & 30M & [62, 0.40, 1.33] \\ \hline
Orion \& Taurus & 829 & -69.8 & 62.9 & 4.04 & 2.38 & 24.7 & 2,856 & 36M & [56, 1.06, 1.90] \\ \hline
Kek \& Kauket & 1040 & -32.3 & 38.6 & 3.25 & 2.06 & 43.2 & 3,461 & 40M & [114, 0.96, 1.68] \\ \hline
Hamilton \& Burr & 941 & -18.0 & 37.7 & 3.26 & 1.17 & 24.7 & 2,882 & 32M & [54, 1.39, 0.57] \\ \hline
Lincoln \& Douglas & 780 & -86.6 & 42.4 & 3.89 & 1.90 & 18.2 & 2,801 & 33M & [60, 1.86, 1.16] \\ \hline
Serena \& Venus$^*$ & 687 & -109.0 & 71.0 & 4.26 & 1.94 & 24.9 & 4,797 & 55M & [159, 0.89, 1.54] \\ \hline
Sonny \& Cher & 966 & -104.9 & 42.0 & 3.69 & 1.05 & 9.7 & 2,290 & 29M & [84, 0.99, 0.84] \\ \hline
Hall \& Oates & 980 & -8.9 & 43.7 & 2.71 & 1.26 & 14.5 & 1,713 & 24M & [64, 1.07, 1.59] \\ \hline
Thelma \& Louise & 832 & -52.4 & 11.0 & 2.36 & 1.30 & 5.3 & 1,693 & 20M & [64, 1.13, 0.46] \\ \hline
Siegfried \& Roy$^*$ & 878 & -68.5 & 57.6 & 4.31 & 1.02 & 11.9 & 5,087 & 61M & [157, 0.61, 1.09] \\ \hline \hline
Milky Way \& M31  & 770 $\pm$ 80 $^g$  & -109 $\pm$ 9 $^g$ & \textless 52  $^g$ & 3.8 $\pm$ 0.7 $^h$ & 1.26 $^{+0.69}_{-0.24}$ $^i$  & -- & -- & -- & [64, 0.89, 0.45]$^j$ \\ \hline
\end{tabular}
\end{center}
\begin{flushleft}
	\begin{scriptsize}
	\begin{multicols}{2}
	\begin{description}
		\item[$^{a}$] Sum of virial masses: $M_\mathrm{v,1} + M_\mathrm{v,2}$
		\item[$^{b}$] Ratio of virial masses: $M_\mathrm{v,2}/M_\mathrm{v,1}$, where $M_\mathrm{v,2} \ge M_\mathrm{v,1}$ by definition.
		\item[$^{c}$] Bi-spherical volume of the high resolution region at $z=0$ that is uncontaminated by low-resolution particles.  Specifically, $\mathcal{V}_\mathrm{res}$ is defined as the union of the two maximal spheres, centered on each pair, that are uncontaminated.
		\item[$^{d}$] Number of identified haloes  with V$_\mathrm{max} >$ 8 km/s that sit within the high-resolution volume $\mathbf{\mathcal{V}}_\mathrm{res}$.
		\item[$^{e}$] Number of particles in millions (rounded to the nearest million) within the high-resolution volume $\mathbf{\mathcal{V}}_\mathrm{res}$.
		\item[$^{f}$] The value of $\vmax$ for and distances to the largest halo within 1.2 Mpc of either host that is not within 300 kpc of either host.  The distances listed are relative to the larger and smaller of the two hosts, respectively.
		\item[$^g$]  As given in \cite{Marel2012} with 2-$\sigma$ uncertainties quoted.
		\item[$^h$]  In listing this value, we average the timing argument result $M_\mathrm{v,1} + M_\mathrm{v,2} = (4.3 \pm 1.1)\times10^{12} \msun$ from \cite{Marel2012} and the sum of our fiducial $\mvir^{\rm MW}$ and $\mvir^{\rm M31}$  values listed in Table~\ref{tab:solopairprops}.  Quoted uncertainties are 2-$\sigma$.  
		\item[$^i$] The quoted average and ratio takes into account that the quantity is defined to be larger than unity.  It combines the constraints listed in Table~\ref{tab:solopairprops} and quotes 90\% uncertainties.
		\item[$^j$] We list the most luminous galaxy within $1.2$ Mpc of either the MW or M31 according to 
	\citet{McConnachie2012}:  NGC 6822 with L$_{\rm V} = 1.04 \times 10^8$ L$_{\odot}$ and $M_\star = 8.3 \times 10^7~\msun$.   We let \textbf{D}$_\mathrm{\ell} =$ \textbf{D}$_\mathrm{M31}$  and \textbf{D}$_\mathrm{s} =$  \textbf{D}$_\mathrm{MW}$.  The $\vmax$ listed for NCG 6822 is very rough, and is based on
	assuming the abundance matching prescription described in Section~\ref{ssec:mstarfuncs} and the $\vmax-M$ relation in Figure~\ref{fig:Mpeakrelation}.  Note that the galaxy IC 1613 is only slightly less luminous than NGC 6822 but is approximately 370 kpc closer to M31 and 300 kpc farther from the MW.  
	\item[$^*$] In order to avoid bias, these pairs are indicated with dashed lines in Figures~\ref{fig:MpDenver},~\ref{fig:MpMpc},~\ref{fig:vmaxdenver},~and~\ref{fig:vmaxmpc} and have been excluded from Figures~\ref{fig:Vradhist}, \ref{fig:fbacksplash}, \ref{fig:Vhist48}, \ref{fig:MstarFuncs}, \ref{fig:Mgasfunc}, and \ref{fig:Vhistgas} because they have large companions at $\sim 1$ Mpc distances.
	\end{description}
	\end{multicols}
	\end{scriptsize}
\end{flushleft}
\caption{Properties of the 12 ELVIS pairs together with associated properties of the MW/M31 pair, where appropriate. Detailed information about the individual haloes that make up these pairs is given in
	Table~\ref{tab:solopairprops}, where they are referred to by the same names
	used in Column 1.}
\label{tab:pairprops}
\end{table*}

\subsection{General Properties of the ELVIS haloes}
\label{subsec:properties}

Table ~\ref{tab:pairprops} summarizes the names and some properties of the ELVIS
pairs at $z=0$ (along with comparative information for the Milky Way and M31,
where appropriate).  We include the physical separation between halo centers,
their relative radial and tangential velocities,\footnote{The kinematics of our
  pairs as listed in Table~\ref{tab:pairprops} are consistent with those found for
  a larger number of pairs in simulations by \citet{Forero-Romero2013}.} as well
as their virial masses and virial mass ratios.  Column 7 lists a conservative
estimate of the high-resolution simulation volume $\mathcal{V}_{\rm res}$,
defined as the union of the two maximal spheres, centred on each pair, that is
uncontaminated by any lower resolution particles.  Columns 8 and 9 list the
overall number of haloes (above our completeness limit of $\vmax > 8 ~\kms$) and
number of simulation particles contained within the volume $\mathcal{V}_{\rm
  res}$.  The final column lists the $\vmax$ value of and distances to the
largest halo within 1.2 Mpc of either host (but outside of the 300 kpc virial
region), which serve as an indication of the larger-scale environment.  Note
that the virial volumes of Hera and Zeus slightly overlap; however, only a
single subhalo is identified in that overlapping volume, so the effect on
subsequent results is negligible.

Two of the pairs -- Siegfried \& Roy and Serena \& Venus -- have a particularly
large halo ($\vmax =157, \, 159\,\kms$) within 1.2 Mpc of one of the hosts.
This may seem contrary to our isolation criteria, but in both cases this third
halo is less massive than either of the paired hosts.  Nevertheless, the
presence of the massive companions may render these pairs less than ideal
comparison sets for the real Local Group.  In all figures below that make
predictions for the overall count of galaxies expected within $\sim$ Mpc scales,
we either remove these two pairs entirely, or show the affected systems with
dashed lines.

\begin{figure*}  
\centering
\includegraphics[width=0.975\textwidth]{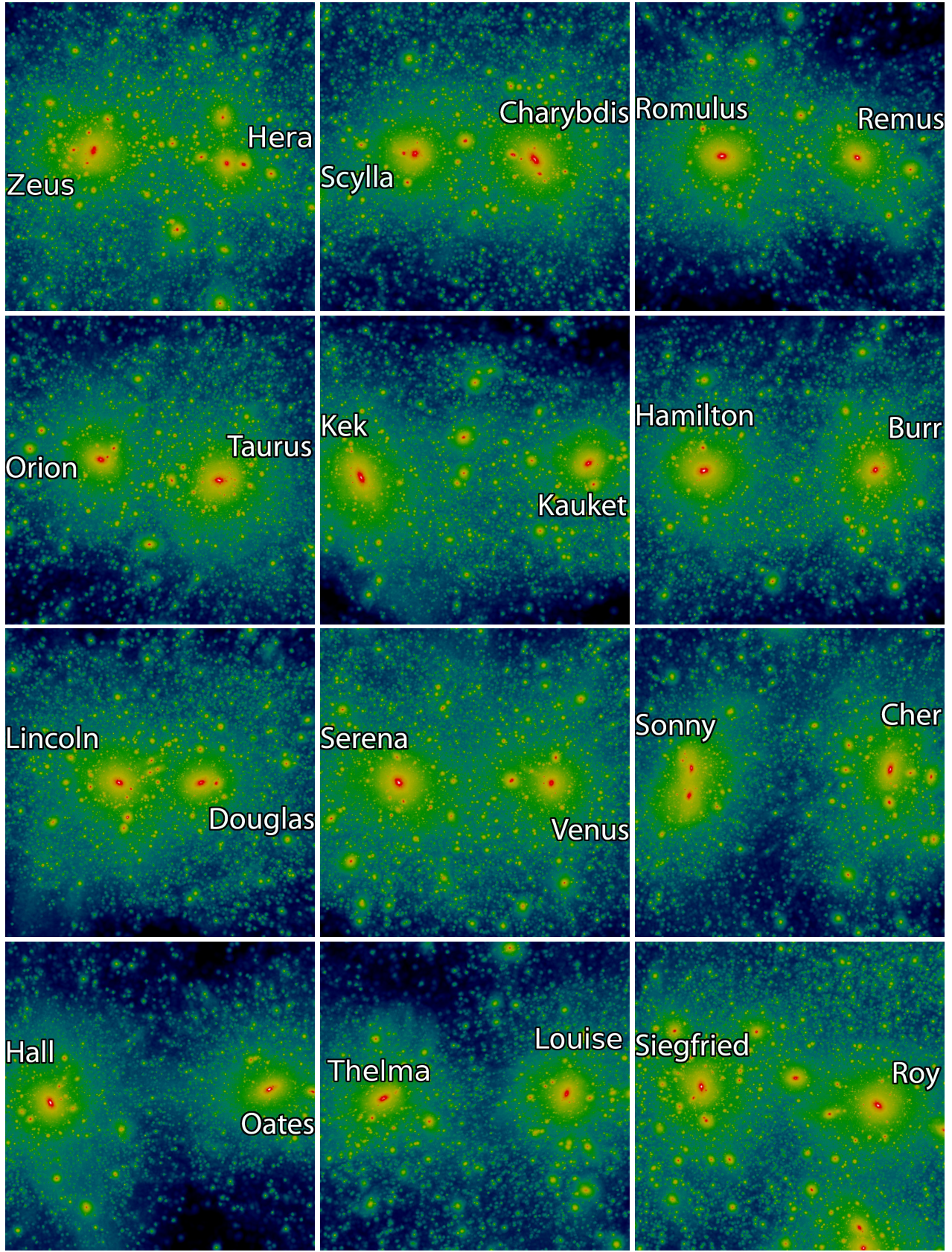}
\caption{Visualizations of the ELVIS pairs, shown in cubes 1.5 Mpc on a side,
  each centred on the mean centre of the pair with names given.}
\label{fig:vizes}
\end{figure*}

Figure~\ref{fig:vizes} shows visualizations of our LG analogs coloured by the
locally smoothed density; each box renders a cube 1.5~Mpc on a side centred on
the midpoint of the two hosts.  Pair names are indicated and the visualizations
are rotated such that the pair is aligned with the horizontal axis, though not
necessarily with an orientation that maximizes the apparent separation.  Each of
these images is fully resolved without contamination from low-resolution
particles, so the shape of the density fields represented are accurate.  There
are a number of features of interest in these images.  For example, it is
readily apparent that Sonny (of Sonny \& Cher in the bottom row) is undergoing a
major merger.  It has a subhalo of $\vmax =115~ \kms$, which is comparable to
the host halo's $\vmax = 180~ \kms$ --- not unlike M33 paired with M31.  Also,
the third massive object near Siegfried \& Roy (as discussed above) is evident
in the bottom-right panel.  As we will discuss below, Zeus \& Hera (upper left)
furnishes a particularly good match to the LG in many observational
comparisons --- the $89~\kms$ subhalo of Hera is shown on the right.

\begin{table*}		
\begin{center}
\setlength{\tabcolsep}{3pt}
\begin{tabular}{|l|c|c|c|c|c|c|c|c|c|c|c|c|}
\hline
\textbf{Halo} & \textbf{M}$_\mathrm{v}$ & \textbf{V}$_\mathrm{v}$ & \textbf{V}$_\mathrm{max}$ & \textbf{R}$_\mathrm{v}$ & \textbf{c}$_{_{\rm -2}}$ $^a$ & \textbf{z}$_{0.5}$ $^{b}$ & \textbf{N}$_\mathrm{haloes}$ $^c$ & \textbf{N}$_\mathrm{haloes}$ $^d$ & {\tiny{Max}} \textbf{V}$_\mathrm{max}$ $^e$ & \textbf{R}$_\mathrm{res}$ $^f$ & \textbf{N}$_\mathrm{p}$ $^g$ & \textbf{N}$_\mathrm{haloes}$ $^h$ \smallskip \\  
 & ($\mathrm{10^{12}M_\odot}$)  & (km/s) &  (km/s) & (kpc) &  &  &  ($<\rvir$)   &  ($<300$)     &   (km/s) &  (Mpc) & ($<R_{\rm res}$) & ($<R_{\rm res}$) \\ \hline\hline
Hera & 1.30 & 140 & 159 & 285 & 7.9 & 0.79 &  397 & 435 & 89 & 1.33 & 39M & 3,348 \\ \hline
Zeus & 2.67 & 178 & 203 & 362 & 5.6 & 1.08 &  1,029 & 880 & 70 & 1.92 & 44M & 3,955 \\ \hline
Scylla & 1.62 & 151 & 179 & 306 & 6.4 & 1.24  &  577 & 567 & 84 & 1.28 & 36M & 3,171 \\ \hline
Charybdis & 2.35 & 171 & 208 & 346 & 7.6 & 0.89  &  896 & 785 & 77 & 1.91 & 47M & 4,368 \\ \hline
Romulus & 1.90 & 159 & 197 & 323 & 9.6 & 1.57  &  623 & 579 & 54 & 1.76 & 30M & 2,427 \\ \hline
Remus & 1.24 & 138 & 177 & 280 & 12.3 & 1.53  &  440 & 463 & 40 & 1.42 & 26M & 2,027 \\ \hline
Orion & 2.84 & 182 & 225 & 369 & 5.3 & 1.61  &  955 & 775 & 47 & 1.60 & 35M & 2,784 \\ \hline
Taurus & 1.19 & 136 & 169 & 276 & 10.9 & 1.08  &  383 & 419 & 61 & 1.21 & 31M & 2,321 \\ \hline
Kek & 2.19 & 167 & 205 & 338 & 13.7 & 0.64 &  685 & 609 & 43 & 1.87 & 39M & 3,333 \\ \hline
Kauket & 1.06 & 131 & 157 & 266 & 9.6 & 1.10  &  388 & 426 & 64 & 1.56 & 32M & 2,687 \\ \hline
Hamilton & 1.76 & 155 & 197 & 315 & 9.9 & 1.47  &  582 & 560 & 62 & 1.39 & 28M & 2,494 \\ \hline
Burr & 1.50 & 147 & 173 & 299 & 10.6 & 1.18  &  613 & 615 & 39 & 1.48 & 29M & 2,529 \\ \hline
Lincoln & 2.55 & 176 & 199 & 356 & 8.4 & 1.36  &  941 & 780 & 75 & 1.27 & 31M & 2,559 \\ \hline
Douglas & 1.34 & 142 & 169 & 287 & 9.6 & 0.99  &  412 & 430 & 89 & 1.32 & 31M & 2,558 \\ \hline
Serena & 2.81 & 181 & 222 & 368 & 14.4 & 1.77  &  911 & 743 & 61 & 1.48 & 51M & 4,418 \\ \hline
Venus & 1.45 & 145 & 156 & 295 & 1.8 & 0.98 &  612 & 623 & 83 & 1.39 & 45M & 3,879 \\ \hline
Sonny & 1.89 & 159 & 180 & 322 & 2.4 & 0.30  &  664 & 637 & 115 & 0.97 & 20M & 1,480 \\ \hline
Cher & 1.80 & 156 & 171 & 317 & 11.0 & 0.66  &  580 & 552 & 81 & 1.12 & 27M & 2,082 \\ \hline
Hall & 1.52 & 148 & 180 & 299 & 10.3 & 1.04  &  437 & 438 & 50 & 1.35 & 23M & 1,560 \\ \hline
Oates & 1.20 & 136 & 167 & 277 & 8.4 & 0.62  &  317 & 346 & 76 & 1.01 & 17M & 1,085 \\ \hline
Thelma & 1.34 & 141 & 169 & 287 & 7.1 & 1.44 & 421 & 438 & 48 & 0.91 & 18M & 1,379 \\ \hline
Louise & 1.03 & 130 & 157 & 263 & 17.0 & 1.61 & 357 & 407 & 54 & 0.80 & 11M & 928 \\ \hline
Siegfried & 2.17 & 166 & 195 & 337 & 6.5 & 0.67  &  827 & 734 & 62 & 1.09 & 46M & 3,674 \\ \hline
Roy & 2.14 & 166 & 194 & 336 & 11.1 & 1.14  &  702 & 628 & 64 & 1.15 & 53M & 4,325 \\ \hline
\hline
Milky Way & 1.6 $^{+0.8}_{-0.6}$ $^i$ & 150 $^{+22}_{-22}$  & -- & 304$^{+45}_{-45}$  & -- & -- & -- &  $\ge$27$^j$ & 88$^k$ & -- & -- & -- \\ \hline
M31 & 1.8 $\pm$ 0.65$^l$ &  156 $^{+17}_{-22}$ & -- & 317$^{+35}_{-44}$  & -- & -- & -- & $\ge$32$^j$ & 130$^m$ & -- & -- & -- \\ \hline
\end{tabular}
\end{center}
\begin{flushleft}
	\begin{scriptsize}
	\begin{multicols}{2}
	\begin{description}
	\item[$^{a}$] Halo concentration defined as $c_{-2} \equiv \rvir/r_{-2}$, where $r_{-2}$ is the radius where $\rho r^2$ peaks.  This parameter is equivalent to the NFW concentration for haloes that follow perfect NFW profiles \citep{NFW96}.
	\item[$^{b}$] Formation time proxy defined as the redshift $z$ when the main progenitor mass first equalled $0.5 \, \mvir(z=0)$.
	\item[$^{c}$] Number of subhaloes within $\rvir$ with $\vmax >$ 8 km/s.
	\item[$^{d}$] Number of subhaloes within 300 kpc with $\vmax >$ 8 km/s. 
	\item[$^{e}$] $\vmax$ value of the largest identified subhalo within 300 kpc.
	\item[$^{f}$] The high-resolution radius, defining a sphere centered on the halo within which there is zero contamination from low-resolution particles.  
	\item[$^{g}$] Number of particles in millions (rounded to the nearest million) within the high-resolution radius  $R_{\rm res}$.
	\item[$^{h}$] Number of subhaloes within $R_{\rm res}$ with $\vmax >$ 8 km/s.
	\item[$^i$] Taken from \cite{MBK2013} with 90\% c.l. quoted.
	\item[$^j$] As enumerated in \cite{McConnachie2012}.
	\item[$^k$] The LMC, from \cite{Olsen2011} 
	\item[$^l$] Combining results from \cite{Fardal2013} and \cite{Marel2012} who obtain $\mvir^{M31} = (2.1 \pm 1.0) \times 10^{12}$ and $(1.5 \pm 0.8) \times 10^{12}$, respectively, with 2-$\sigma$ errors quoted.
	\item[$^m$] The Triangulum galaxy (M33), from \cite{Corbelli2003} 
	\end{description}
	\end{multicols}
	\end{scriptsize}
\end{flushleft}
\caption{Properties of the 24 haloes that comprise our Local Group sample, along with
	the properties of the MW and M31, where appropriate. 
	The haloes are listed in the same order as in Table~\ref{tab:pairprops}, and 
	identified by the names in Column 1 of that Table.  All values in this table are 
	relative to the center of the each host; equivalent properties for the isolated
	sample are listed in Table~\ref{tab:isoprops}, where identical names with
	preceding $i$'s may be used to identified mass-matched analogues.  The table is discussed in \S~\ref{sec:sims}.}

\label{tab:solopairprops}
\newpage
\end{table*}

\begin{table*}		
\begin{center}
\setlength{\tabcolsep}{3pt}
\begin{tabular}{|l|c|c|c|c|c|c|c|c|c|c|c|c|}
\hline
\textbf{Halo} & \textbf{M}$_\mathrm{v}$ & \textbf{V}$_\mathrm{v}$ & \textbf{V}$_\mathrm{max}$ & \textbf{R}$_\mathrm{v}$ & \textbf{c}$_{_{\rm -2}}$ $^a$ & \textbf{z}$_{0.5}$ $^{b}$ & \textbf{N}$_\mathrm{haloes}$ $^c$ & \textbf{N}$_\mathrm{haloes}$ $^d$ & {\tiny{Max}} \textbf{V}$_\mathrm{max}$ $^e$ & \textbf{R}$_\mathrm{res}$ $^f$ & \textbf{N}$_\mathrm{p}$ $^g$ & \textbf{N}$_\mathrm{haloes}$ $^h$ \smallskip \\  
 & ($\mathrm{10^{12}M_\odot}$)  & (km/s) &  (km/s) & (kpc) &  &  &  ($<\rvir$)   &  ($<300$)     &   (km/s) &  (Mpc) & ($<R_{\rm res}$) & ($<R_{\rm res}$) \\ \hline\hline
iHera & 1.22 & 137 & 163 & 278 & 7.9 & 0.8 &  420 & 450 & 41 & 1.54 & 17M & 1,348 \\ \hline
iZeus & 2.59 & 176 & 205 & 358 & 5.5 & 1.3 &  925 & 773 & 60 & 1.76 & 27M & 2,312 \\ \hline
iScylla & 1.59 & 150 & 176 & 304 & 9.9 & 0.97 &  437 & 436 & 84 & 1.56 & 20M & 1,500 \\ \hline
iCharybdis & 2.29 & 169 & 207 & 343 & 13.7 & 1.4 &  758 & 643 & 51 & 1.72 & 25M & 2,125 \\ \hline
iRomulus & 1.97 & 161 & 186 & 327 & 11.3 & 0.88 &  792 & 734 & 75 & 1.89 & 21M & 1,899 \\ \hline
iRemus & 1.31 & 141 & 166 & 285 & 8.0 & 0.91 &  494 & 515 & 54 & 1.40 & 14M & 1,261 \\ \hline
iOrion & 2.84 & 182 & 218 & 369 & 4.9 & 1.64 &  1,179 & 1,015 & 54 & 2.06 & 37M & 3,315 \\ \hline
iTaurus & 1.23 & 138 & 165 & 279 & 10.4 & 1.36 &  453 & 481 & 46 & 1.75 & 14M & 1,315 \\ \hline
iKek & 2.41 & 172 & 204 & 349 & 5.5 & 0.74 &  705 & 618 & 71 & 1.63 & 27M & 2,267 \\ \hline
iKauket$^\dagger$ & 1.02 & 129 & 157 & 262 & 11.1 & 0.97 &  278 & 327 & 39 & 2.12 & 21M & 1,730 \\ \hline
iHamilton & 1.86 & 158 & 203 & 321 & 14.2 & 2.11 &  566 & 523 & 57 & 1.55 & 17M & 1,309 \\ \hline
iBurr & 1.56 & 149 & 179 & 302 & 13.6 & 0.75 &  548 & 540 & 66 & 1.61 & 15M & 1,279 \\ \hline
iLincoln & 2.62 & 177 & 213 & 359 & 13.8 & 0.89 &  813 & 702 & 83 & 1.35 & 27M & 2,017 \\ \hline
iDouglas & 1.30 & 140 & 180 & 285 & 16.1 & 1.76 &  375 & 383 & 49 & 1.93 & 15M & 1,107 \\ \hline
iSerena & 2.67 & 178 & 212 & 361 & 11.4 & 1.15 &  952 & 817 & 81 & 1.66 & 26M & 2,218 \\ \hline
iVenus & 1.39 & 143 & 179 & 291 & 14.3 & 1.41 &  461 & 483 & 46 & 2.15 & 32M & 2,684 \\ \hline
iSonny & 1.68 & 153 & 187 & 310 & 4.5 & 0.69 &  647 & 632 & 117 & 2.01 & 20M & 1,877 \\ \hline
iCher & 1.92 & 160 & 170 & 324 & 6.4 & 0.6 &  701 & 660 & 63 & 2.23 & 22M & 1,888 \\ \hline
iHall$^\diamond$ & 1.71 & 148 & 172 & 300 & 6.0 & 1.13 &  528 & 528 & 92 & 1.59 & 16M & 1,264 \\ \hline
iOates & 1.20 & 136 & 157 & 277 & 8.4 & 0.72 &  444 & 478 & 78 & 1.58 & 13M & 1,068 \\ \hline
iThelma & 1.39 & 143 & 188 & 291 & 9.6 & 1.56 & 407 & 421 & 37 & 1.95 & 14M & 1,043 \\ \hline
iLouise & 1.01 & 129 & 155 & 261 & 8.4 & 1.22 & 378 & 414 & 49 & 2.41 & 14M & 1,253 \\ \hline
iSiegfried & 2.40 & 172 & 211 & 349 & 11.1 & 1.42 &  733 & 643 & 55 & 1.36 & 21M & 1,589 \\ \hline
iRoy & 2.26 & 169 & 205 & 342 & 3.9 & 1.11 &  844 & 769 & 103 & 1.75 & 22M & 1,850 \\ \hline
\hline
iScilla HiRes  & 1.61 & 150 & 175 & 305 & 9.5 & 0.95  & 419             & 413      & 87 & 1.54 & 155M & 1,491  \\    
		    &         &         &        &       &    &     & (3,824)$^\star$ & (3,770)$^\star$ &       &          &          &  (12,509)$^\star$ \\ \hline
iKauket HiRes $^{\ddagger}$ & 1.03 & 130 & 158 & 263 & 11.8 & 1.0 & 277 & 324 & 38 & 0.4 & 56M & 446  \\  
	    &         &         &        &       &    &     & (2,279)$^\star$ & (2,620)$^\star$ &       &          &          & (3,493)$^\star$ \\ \hline            
iHall HiRes  & 1.67 & 152 & 167 & 309 & 5.8 & 1.07  & 608 & 592 & 93 & 1.59 & 125M & 1,286  \\  
	    &         &         &        &       &    &     & (5,266)$^\star$ & (5,114)$^\star$ &       &          &          &  (11,176)$^\star$ \\ \hline 
\hline               
\end{tabular}
\end{center}
\begin{flushleft}
	\begin{scriptsize}
	\begin{multicols}{2}
	\begin{description}
	\item[$^{a}$] Halo concentration defined as $c_{-2} \equiv \rvir/r_{-2}$, where $r_{-2}$ is the radius where $\rho r^2$ peaks.  This parameter is equivalent to the NFW concentration for haloes that follow perfect NFW profiles \citep{NFW96}.
	\item[$^{b}$] Formation time proxy defined as the redshift $z$ when the main progenitor mass first equaled $0.5 \, \mvir(z=0)$.
	\item[$^{c}$] Number of subhaloes within $\rvir$ with $\vmax >$ 8 km/s.
	\item[$^{d}$] Number of subhaloes within 300 kpc with $\vmax >$ 8 km/s. 
	\item[$^{e}$] $\vmax$ value of the largest identified subhalo within $300$ kpc. 
	\item[$^{f}$] The high-resolution radius, defining a sphere centered on the halo within which there is zero contamination from low-resolution particles.  
	\item[$^{g}$] Number of particles in millions (rounded to the nearest million) within the high-resolution radius  $R_{\rm res}$.
	\item[$^{h}$] Number of subhaloes within $R_{\rm res}$ with $\vmax >$ 8 km/s.
	\item[$^\dagger$] iKauket was initialized at $z = 250$ for both the fiducial and HiRes runs.
	\item[$^\diamond$] Differences in the phase of subhalo orbits between this run and the HiRes equivalent result in the largest subhalo ($\vmax = 93\kms$) being located just beyond $\rvir$ at the fiducial resolution.  To show convergence with the HiRes run, we include the mass of that halo in the virial mass of iHall and list it in Column 10, though it is just beyond $300$ kpc.  The uncorrected substructure counts are also divergent, but the number of objects within $400$ kpc agrees within 5\%.  The uncorrected mass ($1.53\times10^{12}\msun$) also agrees with the paired halo, Hall, to within 1\%. 
	\item[$^\star$] Values in parentheses correspond to subhalo counts down to $\vmax >$ 4 km/s, the estimated completeness limit of the HiRes simulations.
	\item[$^{\ddagger}$] This halo was initialized with a smaller Lagrange volume of high-resolution particles than the rest, which is why it has an anomalously small high-resolution radius.
	\end{description}
	\end{multicols}
	\end{scriptsize}
\end{flushleft}
\caption{Properties of the 24 isolated haloes that are mass matched to 
		the haloes in our Local Group analogues.  The name identifies the paired 
		halo with a nearly identical mass, the properties of which are listed in 
		Table~\ref{tab:solopairprops}, and the preceding $i$ indicates an isolated 
		analogue.  Columns are identical to those in Table~\ref{tab:solopairprops}.  The last three rows
		correspond to the HiRes simulations of three haloes.}
\newpage

\label{tab:isoprops}
\end{table*}

We list the properties of the individual haloes that comprise each pair in
Table~\ref{tab:solopairprops} along with comparative information for the MW 
and M31, when appropriate.  A similar list for the isolated mass-matched
analogs is given in Table~\ref{tab:isoprops}.  In each table,
Columns~2~through~5 list $\mvir$, $\vvir$, $\vmax$, and $\rvir$, respectively.
Column~6 gives a measure of the halo concentration, $c_{-2} \equiv
\rvir/r_{-2}$, where $r_{-2}$ is the radius where $\rho r^2$ peaks [equivalent
to the concentration parameter for haloes that follow \citet[NFW]{NFW96}
profiles].  Column~7 provides a measure of the halo formation redshift,
$z_{0.5}$, defined when the main progenitor first obtains half its current mass.
Columns~8~and~9 list the number of $\vmax > 8 ~\kms$ subhaloes within $\rvir$ and
300~kpc, respectively, and column~10 lists the $\vmax$ of the largest subhalo
within 300~kpc.  Column~11 gives $R_{\rm res}$, the radius of the largest sphere
within which there are no low-resolution particles (an indication of the
high-resolution volume size).  Columns~12 and 13 list the number of particles
(in millions, rounded to the nearest million) and number of identified haloes
(with $\vmax > 8 ~\kms$) within $R_{\rm res}$ for each halo.

Note that in what follows we will occasionally present results for a region we
define as the Local Volume -- the union of two spheres of radius 1.2 Mpc
centred on each host.  As can be seen from Column~11 of
Table~\ref{tab:solopairprops}, four of our pairs are technically contaminated in
this region (Sonny \& Cher, Hall \& Oats, Thelma \& Louise, and Siegfried \&
Roy).  However the mass fraction of low-resolution particles in the effected
volumes is minimal ($0.01$, $0.007$, $0.06$, and $0.0008$ per cent
respectively) so the practical effects on our results should be negligible
\citep[see, e.g.][]{Onorbe2013}.

Before devoting the next section to a detailed comparison of paired versus unpaired
hosts, we mention that we find no statistical difference in the $c_{-2}$ and
$z_{0.5}$ distributions between the two sets.  Though two of our haloes (Serena
and Sonny) that happen to be members of pairs have anomalously low $c_{-2}$
values, we suspect that in Sonny's case this is a result of an ongoing major
merger.  The median formation redshifts for our paired and unpaired samples are
both $z_{0.5} \simeq 1.1$, with no indication that paired halo formation times
correlate.

The lack of difference in the $z_{0.5}$ distribution between the two samples is
consistent with the comparison made by \citet{Forero-Romero2011} using similarly
paired haloes found in the Bolshoi simulations.  These authors point out that
the three LG-like pairs identified in the constrained CLUES simulations have
anomalously early formation times, all three with half-mass formation times
$z_{0.5} \gtrsim 1.5$.  Three of our 12 paired systems are similarly
early-forming (Romulus \& Remus, each with $z_{0.5} \simeq 1.6$), Orion \&
Taurus (with $z_{0.5} = 1.6$ and 1.3, respectively), and Thelma \& Louise (with
$z_{0.5} = 1.4$ and $1.6$).

\section{Paired \lowercase{versus} Isolated Galaxies}
\label{sec:LGvIso}

\begin{figure}		
  \centering
 \includegraphics[width=0.49\textwidth]{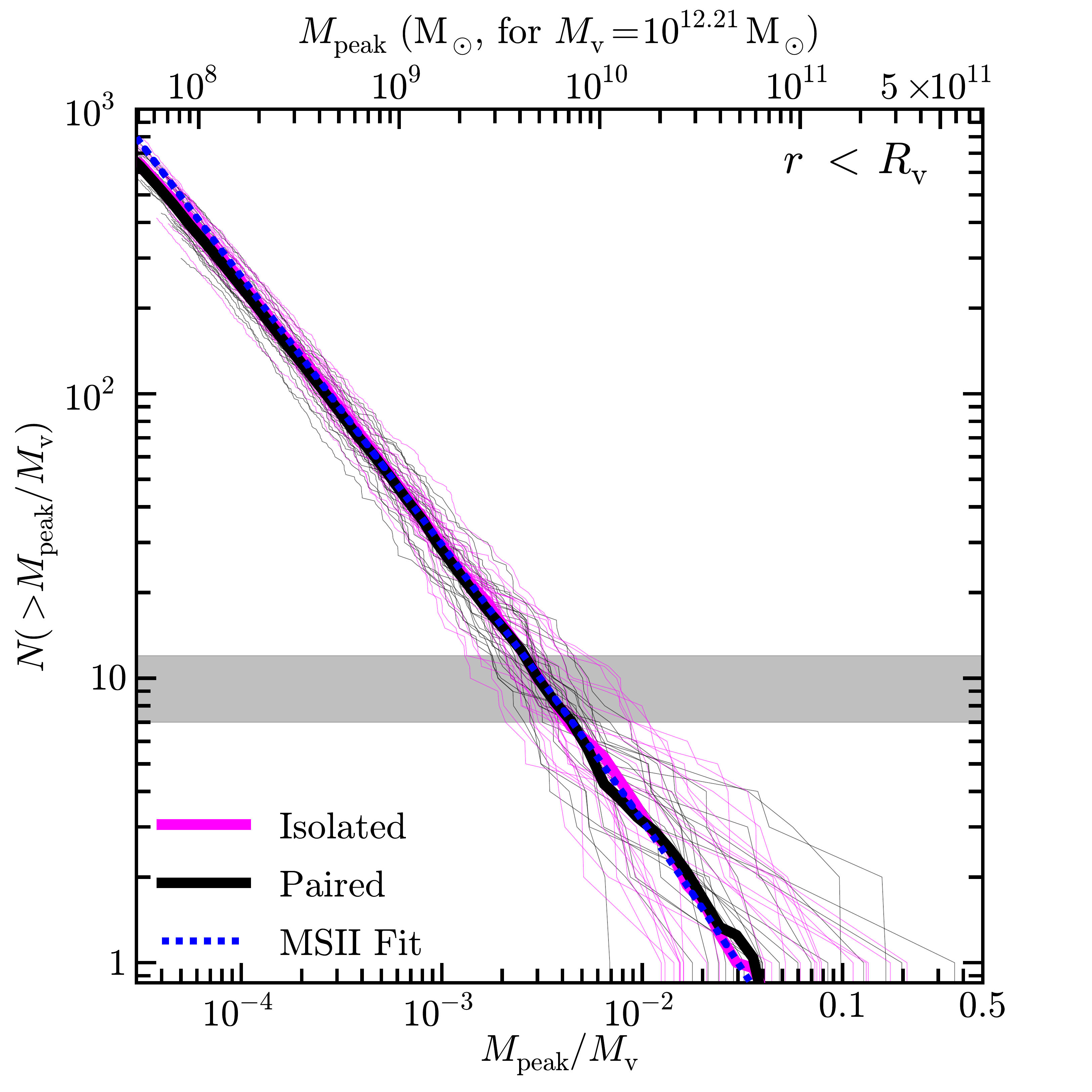}
 \caption{Cumulative subhalo peak mass function ($\mpeak$) normalized by host
   $\mvir$ for each isolated (thin magenta lines) and paired (thin black lines)
   host.  All objects within $\rvir$ are plotted.  The average for each
   population is shown by the thick lines of corresponding colour.
   Statistically, the mass functions for paired and isolated hosts are
   indistinguishable, though the halo-to-halo scatter is large.  The upper axis
   is scaled to the subhalo $\mpeak$ values assuming a host virial mass of
   $\mvir = 1.6\times10^{12}\msun$, which is our fiducial MW mass.  Thin lines
   are truncated at $\mpeak = 6 \times 10^7 \msun$, the completeness limit of
   our simulation catalogs.  The grey band shows the range in number of
   satellites around the MW and M31 with stellar masses above
   $10^6\msun$; from this band, one can see that such galaxies would be expected
   to form in haloes more massive than $\mpeak \simeq 3 \times 10^{-3} \mvir
   \simeq 5 \times10^9\msun$.  }
  \label{fig:MpRvir}
\end{figure}

\begin{figure}		
  \centering
  \includegraphics[width=0.49\textwidth]{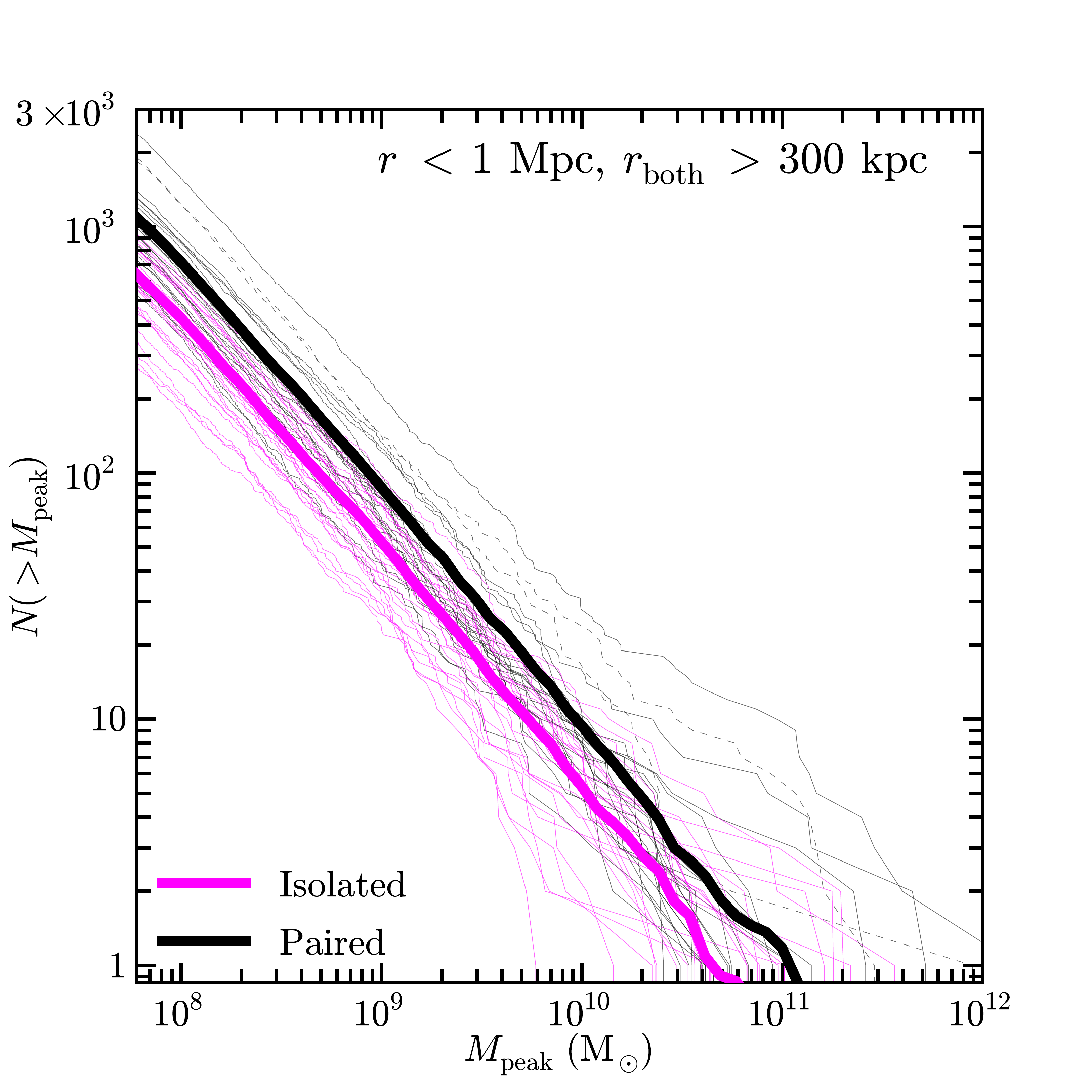}
  \caption{Cumulative counts, as a function of $\mpeak$, for haloes that are
    between 300~kpc and 1~Mpc of any host.  The paired population (black) has an
    amplitude that is approximately 80\% larger at fixed $\mpeak$ than that of the
    isolated analogs (magenta).  The environment around a LG pair thus differs noticeably 
    from that of an isolated MW-size halo, even though such differences are not 
    manifest within the virial radius (Figure~\ref{fig:MpRvir}).}
  \label{fig:MpDenver}
\end{figure}

\subsection{Halo abundances}
\label{subsec:abundance}
 \begin{figure}		
	\centering
	\includegraphics[width=0.49\textwidth]{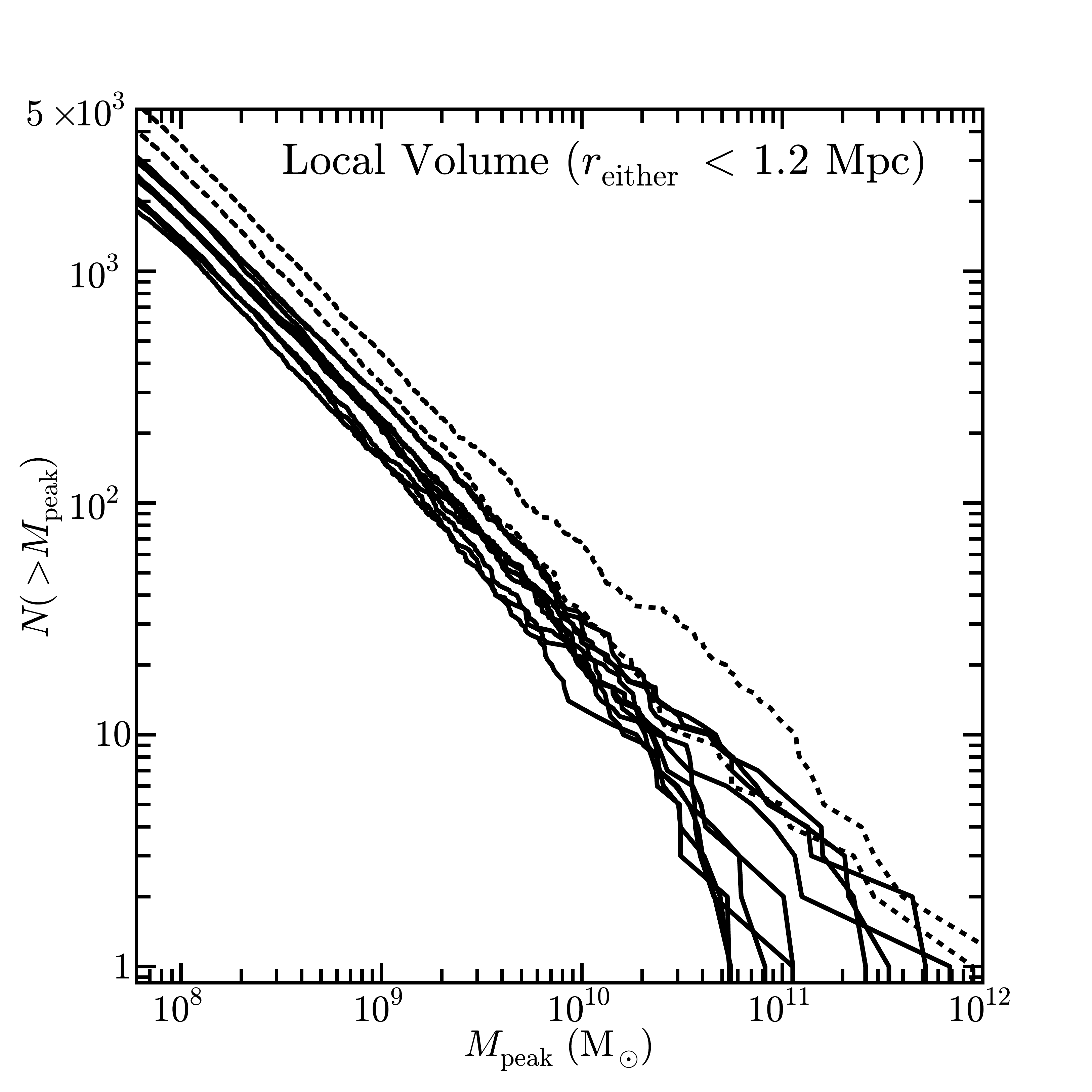}
	\caption{ The $\mpeak$ functions around the LG pairs; each line
          represents a pair of giants and includes all haloes (excluding the MW
          and M31 analogs) within $1.2~\mpc$ of either host, which we define as
          the Local Volume.  Two pairs contain a third large system within the
          volume and are thus shown as dashed lines.  We predict $\sim2000 -
          3000$ objects with $\mpeak > 6\times 10^7\msun$ within the region.}
\label{fig:MpMpc}
\end{figure}

We begin by examining the abundance of dark matter structures,
characterized by their $\mpeak$ values, within various radial boundaries; counts
as a function of $\vmax$ are presented in Appendix~B.  

Figure~\ref{fig:MpRvir} shows the cumulative $\mpeak$ functions for subhaloes
within $\rvir$, normalized to the host halo virial mass $\mvir$, for each of the
48 hosts in ELVIS.  Isolated hosts are shown as thin magenta lines and the
paired hosts are plotted in black.  The two distributions clearly overlap.  The thick lines
denote the mean cumulative count at fixed $\mpeak/\mvir$ for the isolated
(magenta) and paired (black) populations.  Both distributions are well fitted at the
low mass end by a power-law:
\begin{equation}
N_{\rm v} (>M_{\rm peak}) = 3.85 \left( \frac{M_{\rm peak}}{0.01\, \mvir}\right)^{-0.9}.
\end{equation}
Within $\rvir$, subhalo counts within isolated and paired haloes in ELVIS are
indistinguishable. Even for high-mass subhaloes, where the intrinsic scatter in
the counts is large, the means agree well.  The blue dashed line, which sits
practically on top of the ELVIS means, shows the mean power-law fit obtained by
\citet{Boylan-Kolchin2010} for subhaloes in a large sample of MW-mass haloes
from the Millennium-II Simulation \citep[MS-II;][]{Boylan-Kolchin2009}.  The same fit also
matches the substructure counts from the Aquarius simulations \citep{Aquarius}
well.  The agreement between our simulations and MS-II/Aquarius is remarkable,
especially given that the cosmology of these older simulations is slightly
different from our adopted values, which are based on more recent constraints.

Broadly speaking, the scatter in subhalo counts among haloes also agrees between
the two samples.  At small masses ($M_{\rm peak} / \mvir \lesssim 10^{-3}$) we
find that the standard deviation divided by the mean approaches $\sigma/\langle
N \rangle \simeq 0.15$, and that the scatter increases towards higher masses,
with $\sigma/\langle N \rangle \simeq 0.4$ at $M_{\rm peak} \simeq 0.01 \mvir$.
This result is consistent with an intrinsic halo-to-halo scatter of $\sim 15\%$
in the abundance of substructure reported elsewhere
\citep{Boylan-Kolchin2010,Busha2011,Wu2013}.

Though we do not plot it, the $z=0$ (bound) mass functions also agree well
within $\rvir$ and are both well fitted by
\begin{equation}
N_{\rm v} (>M) = 1.11 \left( \frac{M}{0.01\, \mvir}\right)^{-0.95},
\end{equation}
though the scatter is slightly larger than in the $\mpeak$ function ($\sigma/\langle
N\rangle\sim0.2$ at small masses).

One take away from this initial result is that predictions for subhalo counts
within the virial radius from previous high resolution simulations that studied
isolated MW-size hosts \citep[e.g.][]{Diemand2008,Kuhlen2008,Aquarius} are not
expected to be significantly different than those for paired haloes like the
MW and M31.

\begin{figure*}  
  \centering
  \includegraphics[width=0.49\textwidth]{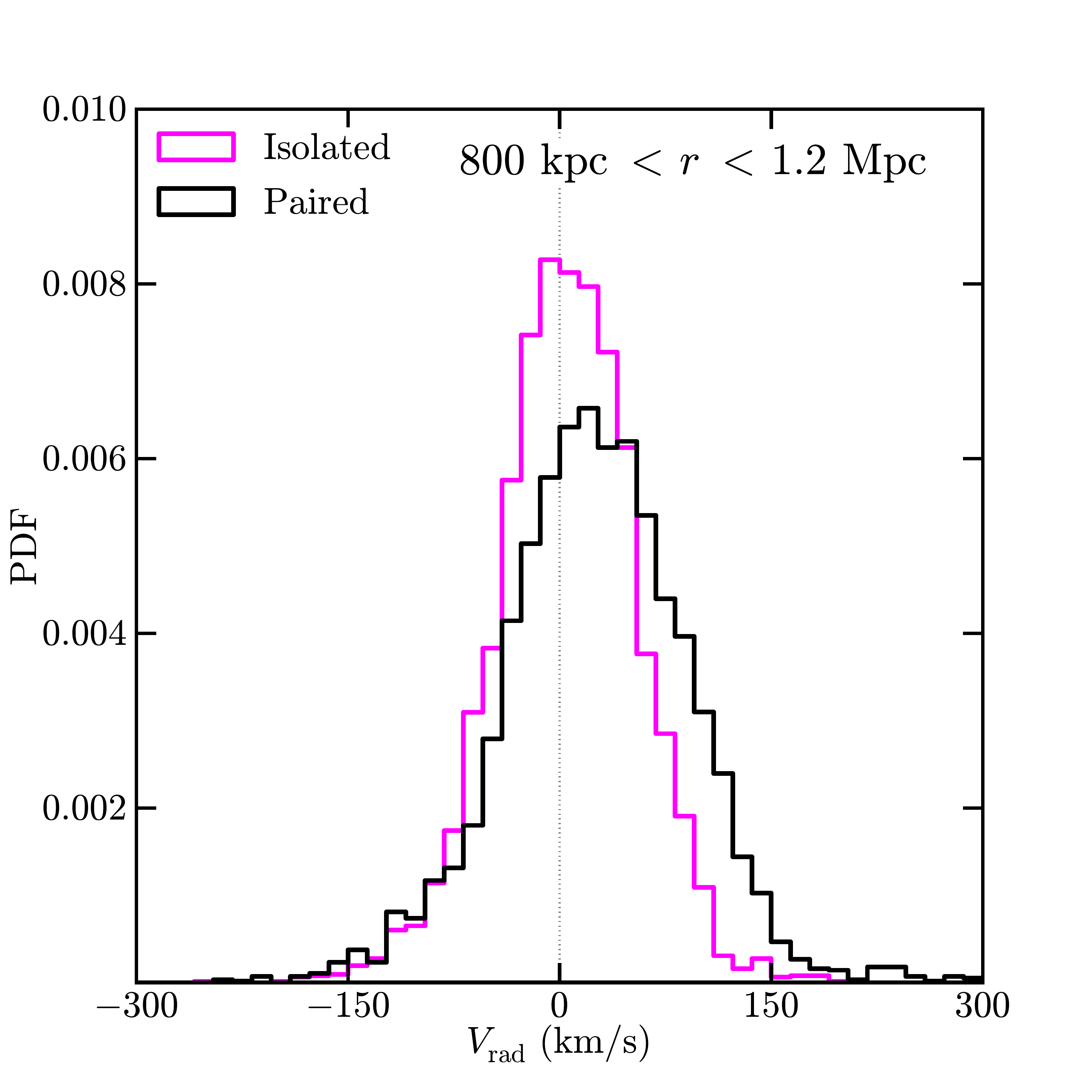}
  \centering
  \includegraphics[width=0.49\textwidth]{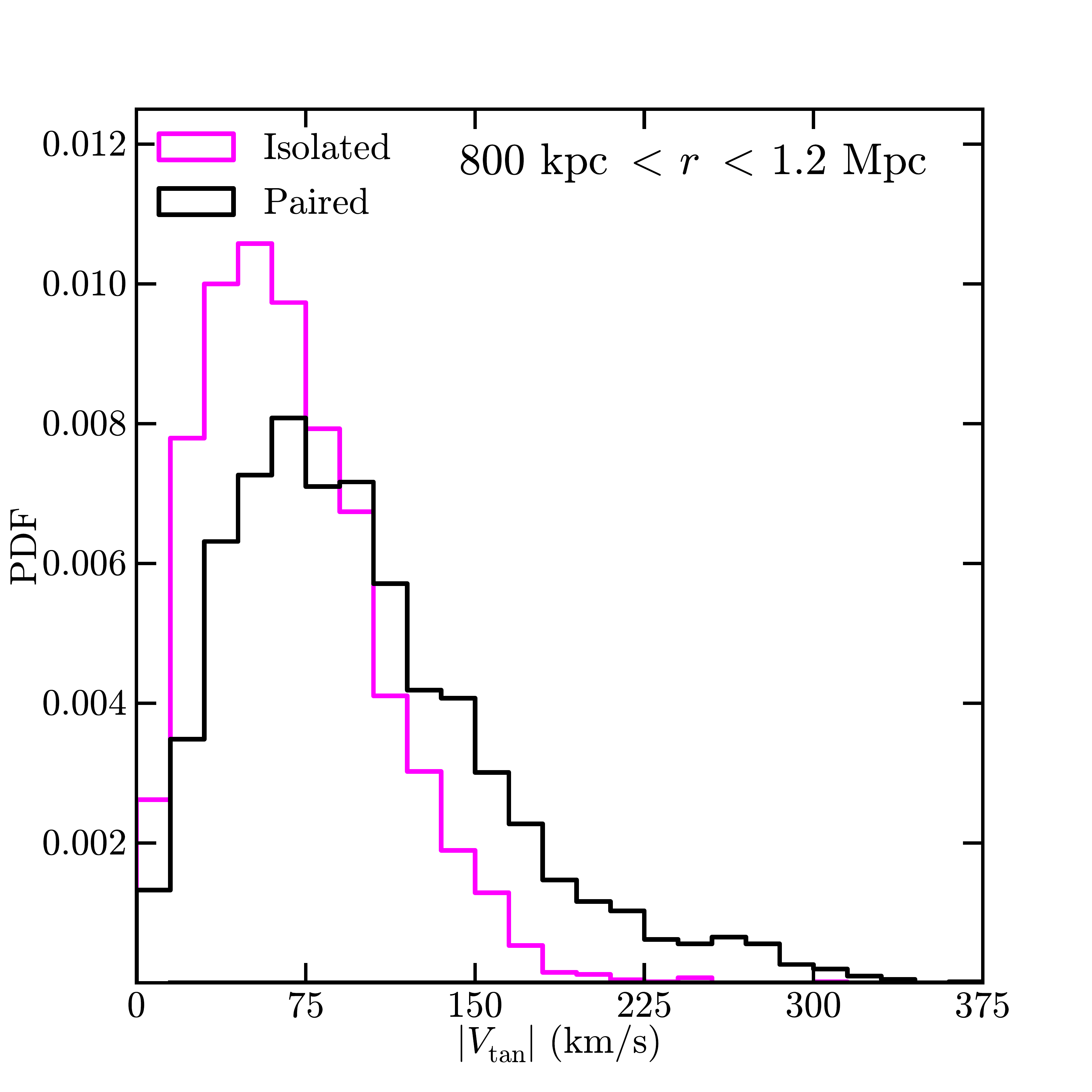} 	
  \caption{Stacked distributions of radial (left panel) and tangential (right
    panel) velocities for haloes around the isolated (magenta) and paired (black)
    haloes at distance of $0.8-1.2\,\mpc$ from the {\em nearest} host.  While the
    distributions for paired and isolated haloes are the same within $\rvir$ (not
    plotted), the differences become pronounced at large radii, with paired
    environments being substantially hotter.  While essentially all haloes
    $\sim1~\mpc$ from isolated MW analogs have $V_{\rm tan} < 200\,\kms$, a
    large fraction around LG analogs have $V_{\rm tan} > 200\,\kms$.  It is also
    apparent that while the radial velocities of haloes at $\sim1$ Mpc distance
    from isolated MW-like hosts are centred on zero, the paired analogs have an
    excess population of outflowing systems.  These outflowing systems include a
    ``backslash'' population that is larger among pairs (see \S~\ref{subsec:backsplash}) and also
    objects that have yet to turn around from the Hubble flow (the zero velocity
    surface is centred on the pair at $\sim 1$ Mpc distance, not the individual
    host).  One broad implication of this Figure is that in order to correctly
    predict the large-scale velocity field around the MW, one must account for
    the presence of M31.}
  \label{fig:Vradhist}
\end{figure*}

At distances beyond the virial radii of either host, the presence of a massive
companion should affect halo abundances.  To compare the counts at large
distances between isolated and paired MW-size haloes, we must avoid the
bias that would be introduced by simply counting all haloes at a given distance,
as many will be subhaloes of the M31 analogue in the paired systems.  We attempt to
remove this bias by defining a region around each host that we call the ``Local
Field'': a spherical shell between 300~kpc and 1~Mpc of the centre of that host,
but excluding the region within 300~kpc of the centre of the other giant. That
is, no subhaloes of either LG giant analogue are included in this region. 

We plot the $\mpeak$ function for these Local Field regions around all the ELVIS haloes
in Figure~\ref{fig:MpDenver}.  The environment surrounding a typical LG-like
halo is richer than that around an isolated system, even when the partner's
subhaloes are removed.  Specifically, the average relations are again well fitted at
the low mass end by power laws, but the normalization for the paired sample is
about 80\% higher than that of the isolated sample:
\begin{equation}
N_{0.3-1}(>M_{\rm peak})=N_0\left(\frac{M_{\rm peak}}{10^{10}\msun}\right)^{-0.9} \,,
\end{equation}
with $N_0=6.4$ for the isolated sample and $N_0=11$ for the paired sample.  The
dashed lines in Figure~\ref{fig:MpDenver} indicate the two haloes that have a
large companion within the $0.3-1$ Mpc region (see \S~\ref{subsec:properties} 
and Table~\ref{tab:pairprops}).  It is possible that these systems are poor 
comparison sets to the Local Group, which appears to lack such a galaxy 
(Table~\ref{tab:pairprops}).  If we remove the dashed lines from the
fit, the normalization for the paired systems becomes $N_0 = 9.2$, which is
$\sim 56 \%$ higher than that for the isolated sample (removing the isolated
counterparts to those haloes also gives a slightly lower normalization $N_0 =
5.9$).  While the distributions show some overlap, the presence of a paired
companion appears to bias the overall large-scale environment to be
substantially richer in small haloes, even when the subhaloes of the paired host
are excluded from the counts.

Figure~\ref{fig:MpMpc} presents total halo number counts as a function of
$\mpeak$ within a bi-spherical volume defined by overlapping spheres of radius
1.2~Mpc of each paired host.  There is one line for every ELVIS pair, thus each
can be regarded as a realization of the LG itself.  The dashed lines
indicate the two pairs that have large companions in the region, possibly making
them less than ideal comparison sets for the LG.  Neglecting those two
systems, the group-to-group scatter in this Local Volume mass function is
remarkably tight, spanning less than a factor of $\sim 2$ over all 10
realizations for masses $\mpeak \lesssim 10^{10}, \msun$.  In total, the best LG
analogs in the ELVIS suite have $2000 - 3000$ haloes with $\mpeak >
6\times10^{7}\,\msun$ in this Local Volume region.  Of course, many of these
small haloes likely contain galaxies that are either devoid of stars entirely, or
too faint to detect with current methods.  We investigate implications for the
number of observable galaxies throughout this region in \S 4.

\subsection{Halo Dynamics}
\label{subsec:dynamics}
We expect that the presence of M31 alters the dynamical structure of the Milky
Way's local environment relative to the environment of an isolated analogue. While
we find that, within $\sim 300$ kpc of the hosts, the paired and isolated
samples have indistinguishable subhalo kinematics, regions beyond this distance
show distinct kinematical differences.

Figure~\ref{fig:Vradhist} shows stacked distributions of radial and tangential
velocities for $\mpeak > 6 \times 10^7 \msun$ haloes having distances between
800~kpc and 1.2~Mpc of a giant.  Note that in these histograms, we compute the
distance to both of the hosts and only use the smaller of the two distances
(i.e., all haloes at distance $r$ from one host are at least that same distance
$r$ from the other host).  Regions surrounding isolated hosts are shown in magenta
while regions around paired systems are in black.  The kinematic distinction is
clear: paired haloes are both kinematically hotter and show an excess of systems
that are outflowing at this radius.  As we discuss in the next subsection, this
enhanced population of outflowing haloes includes a large number of objects that
were once within the virial radius of one of the giants.  This fraction appears
to be higher in paired hosts.  A complication when interpreting the radial
velocity figure is that the zero-velocity/turn-around surface (at $\sim1$ Mpc
distance) for the pairs is centred between the hosts rather than on the main
halo as it is for the isolated analogs.  This means that some fraction of the
haloes in this diagram may not have turned around from the Hubble flow.

In the histograms shown in Figure~\ref{fig:Vradhist}, we have removed haloes
belonging to the pairs Siegfried \& Roy and Serena \& Venus.  As discussed
above, these pairs have a particularly large halo within 1.2 Mpc of one of the
hosts, and therefore may be poor analogs to the real LG.  Including
them only serves to make the overall paired histograms even hotter compared to
the isolated analogs.

 \subsection{Backsplash Halos}
\label{subsec:backsplash}
 
Here we investigate the dynamical histories of each small halo in the vicinity
of our MW analogs at $z=0$, and specifically ask whether a halo has been within
the virial radius of either giant since $z=5$.  If so, then in principle,
environmental effects such as ram pressure, harassment, or strangulation could
have quenched the galaxy it hosts
\citep{Kawata2008,Boselli2008,Grcevich2009,Woo2013,Phillips2013}.  We refer to
previously-interacted objects of this kind as ``backsplash'' haloes (e.g.,
\citealt{Gill2005} and references therein).  \citet{Knebe2011} identified an 
additional population of haloes, which they termed "renegade", that have been a 
member of both the M31 and MW halo analogs.  We reserve a more detailed study
of these interesting objects for a future work~--~for this paper, we combine renegade 
haloes beyond $\rvir$ with all other backsplash haloes and those within $\rvir$ with
all other subhaloes.

\begin{figure}		
  \centering
  \includegraphics[width=0.49\textwidth]{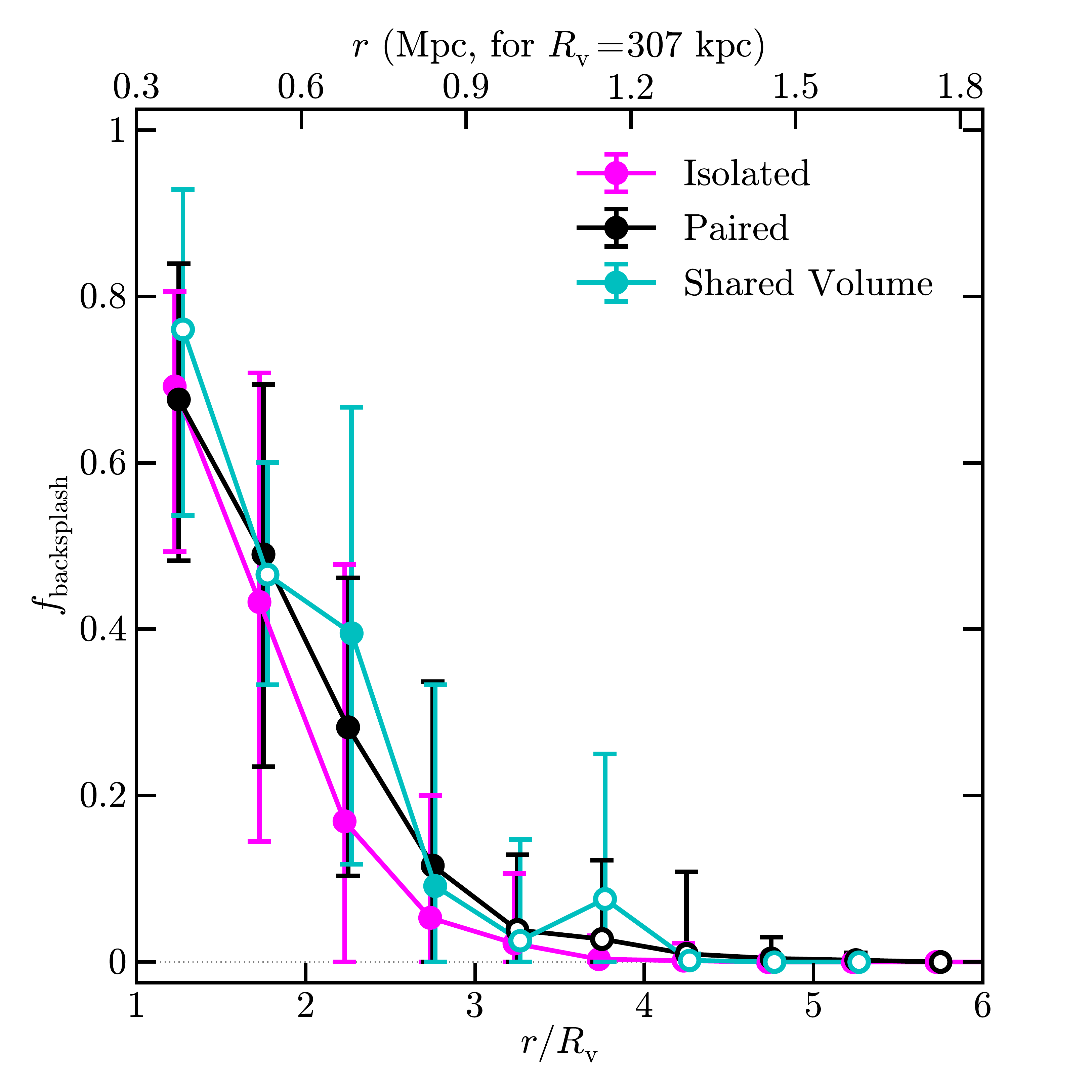}
  \caption{The fraction of $\mpeak > 6\times 10^7\msun$ haloes at $z=0$ that have
    been within $\rvir$ of a MW size host as a function of $r/\rvir$ from the
    centre of each host. The points show the average in each radial bin and the
    error bars denote the full width of the distribution over all hosts.  The
    magenta line corresponds to the isolated sample, and the black line corresponds
    to paired hosts, where the distance is to the nearest of the two giants.  The cyan
    line also counts systems in the paired simulations, but counts only those
    systems that simultaneously meet the radial cut for both hosts.  The most
    likely location for backsplash haloes is in this shared volume and between
    $1$ and $2\,\rvir$ of both hosts (i.e., in between the two haloes rather than on
    one side or the other of the LG pair). }
  \label{fig:fbacksplash}
\end{figure}

Figure~\ref{fig:fbacksplash} presents the differential fraction of haloes that
are backsplash objects as a function of distance from each host in radial bins
of width $\rvir/2$.  Systems around our LG-analogs are shown in
black, where the distance assigned is the minimum of the distances to the two
giants in the group.  As in Figure~\ref{fig:Vradhist}, we have removed haloes
belonging to the two pairs in our sample with large companions at $\sim 1$ Mpc
distance.  The subsample of haloes that meet the radial cut from the centre of
both giants simultaneously are shown in cyan.  The isolated sample is shown in
magenta.  We indicate with open symbols bins where the full halo sample was not
used, either due to contamination at large radii or because there are no haloes
that meet the radial cut in the bin.  The points correspond to the average over
all hosts and the error bars denote the full width of the distribution, measured
system-by-system.

Unsurprisingly, the backsplash fraction is largest at small radii.  In the
regions spanning $1-1.5 \, \rvir$, typically 70\% of haloes have been within
$\rvir$ since $z= 5$, though that number can be as high as 80\% in some cases
\citep[also see][]{Mamon2004,Gill2005}.  The interaction fraction in the
environment of LG-like pairs is systematically higher than in isolated analogs
at large radii, and the overlapping volume (cyan) is particularly rich in
objects that have interacted.  Indeed, the shared region in the real LG may be
the best hunting ground for potential backsplash candidate dwarfs.  Remarkably,
in our LG-analogue systems, the probability that a halo has interacted only drops
to zero at approximately $5\,\rvir\;(\sim 1.5 ~{\rm Mpc})$.  Expressed
cumulatively (rather than differentially), we find that the overall fraction of
backsplash haloes within the 1.2 Mpc Local Volume regions of our paired hosts
ranges from $30 \%$ to $52 \%$.

\begin{figure*} 
\centering
	\includegraphics[width=0.49\textwidth]{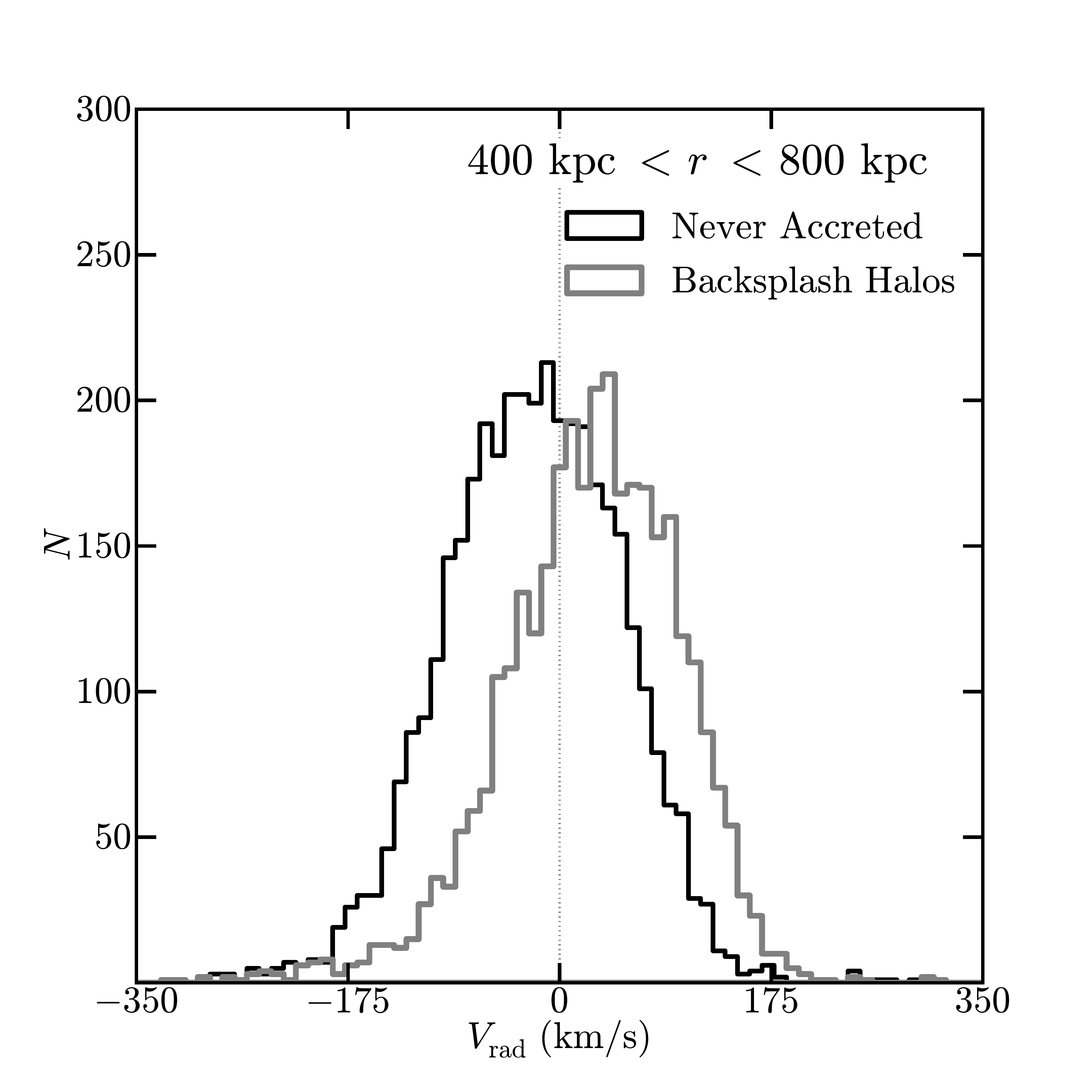}
\centering
	\includegraphics[width=0.49\textwidth]{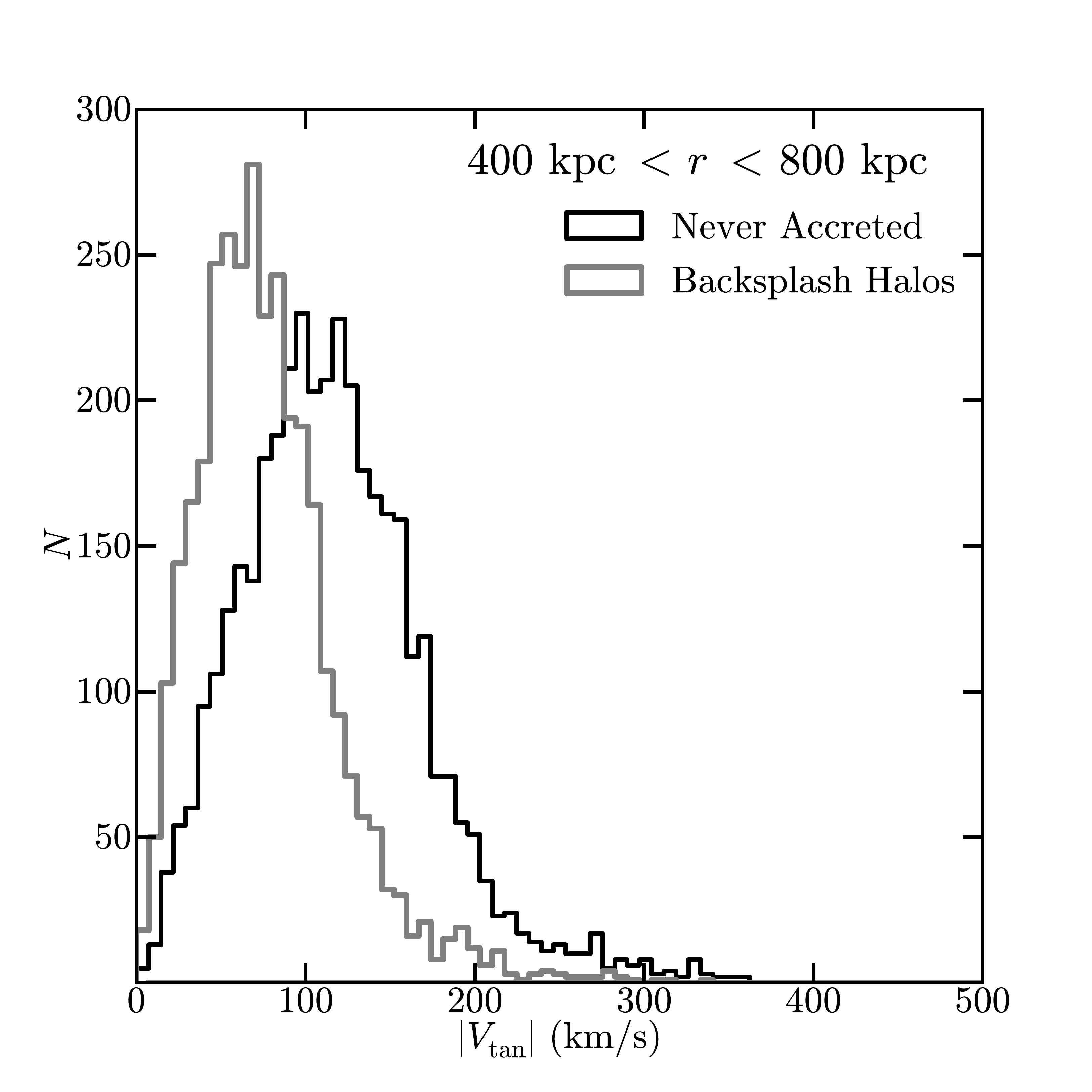}
    \caption{The radial (left) and tangential (right) velocity distributions of
      field haloes in the spherical shell 400--800~kpc from the centre of each
      paired host, truncated in the same manner as
      Figures~\ref{fig:Vradhist}~and~\ref{fig:fbacksplash}.  The black lines
      plot only those haloes that have never been within $\rvir$ and the grey
      lines includes only backsplash haloes.  The latter are comparatively
      outflowing with relatively low tangential velocities.  Note that in this
      figure we have excluded the two pairs in our sample with large companions
      at $\sim 1$ Mpc distance.}
\label{fig:Vhist48}
\end{figure*}

How might backsplash haloes be distinguished observationally throughout the
LG?  In Figure~\ref{fig:Vhist48}, we compare the relative tangential
and radial velocities of backsplash haloes (grey line) in the $r=400 - 800$ kpc radial 
bins to those that have never accreted (black line).  Here we limit ourselves to paired 
hosts.  As in \citet{Teyssier2012}, we find that backsplash haloes tend to be outflowing
from the host that they have interacted with, whereas those that have not yet
accreted are preferentially inflowing.  As the right plot shows, we also expect
backsplash haloes to have low tangential velocities compared to those that have
never been within $\rvir$.  The implication is that backslash systems are more
likely to be on radial orbits and to be on their way out.  At the same time,
there is significant overlap in the distributions and it is difficult to
disentangle the populations on these specific kinematic properties alone.  We
reserve a more thorough analysis of this question for a future paper.

\section{Expectations for the Local Group}
\label{sec:predictions}

As the previous section showed, number counts and velocity distributions 
within $\rvir$ are consistent between isolated and paired MW-size haloes, 
but differences are evident at greater radii.  In this section, we will focus on predictions
in the $\sim1$~Mpc scale environment around the Milky Way and will present
results for the paired sample only.

\subsection{Stellar Mass Functions}
\label{ssec:mstarfuncs}
Although the ELVIS simulations are dissipationless, the abundance matching (AM)
technique \citep{Kravtsov2004,Vale2004, Conroy2006, BehrooziAM,Moster2013} makes
it possible to assign stellar masses to dark matter haloes and convert the halo
mass functions in Figures~\ref{fig:MpRvir}~and~\ref{fig:MpMpc} into reasonably
proxies for stellar mass functions.  The connection between galaxy mass and halo
mass remains highly uncertain at low masses $\mstar \la 10^{8}\,\msun$, however,
as it is difficult to measure luminosity functions over large volumes for dim
galaxies.  In this sense, comparisons to galaxy counts within the LG, where luminosity 
functions are more complete, can help test and refine AM relationships that have been built 
upon cosmological samples.

\begin{figure}		
  \centering
  \includegraphics[width=0.49\textwidth]{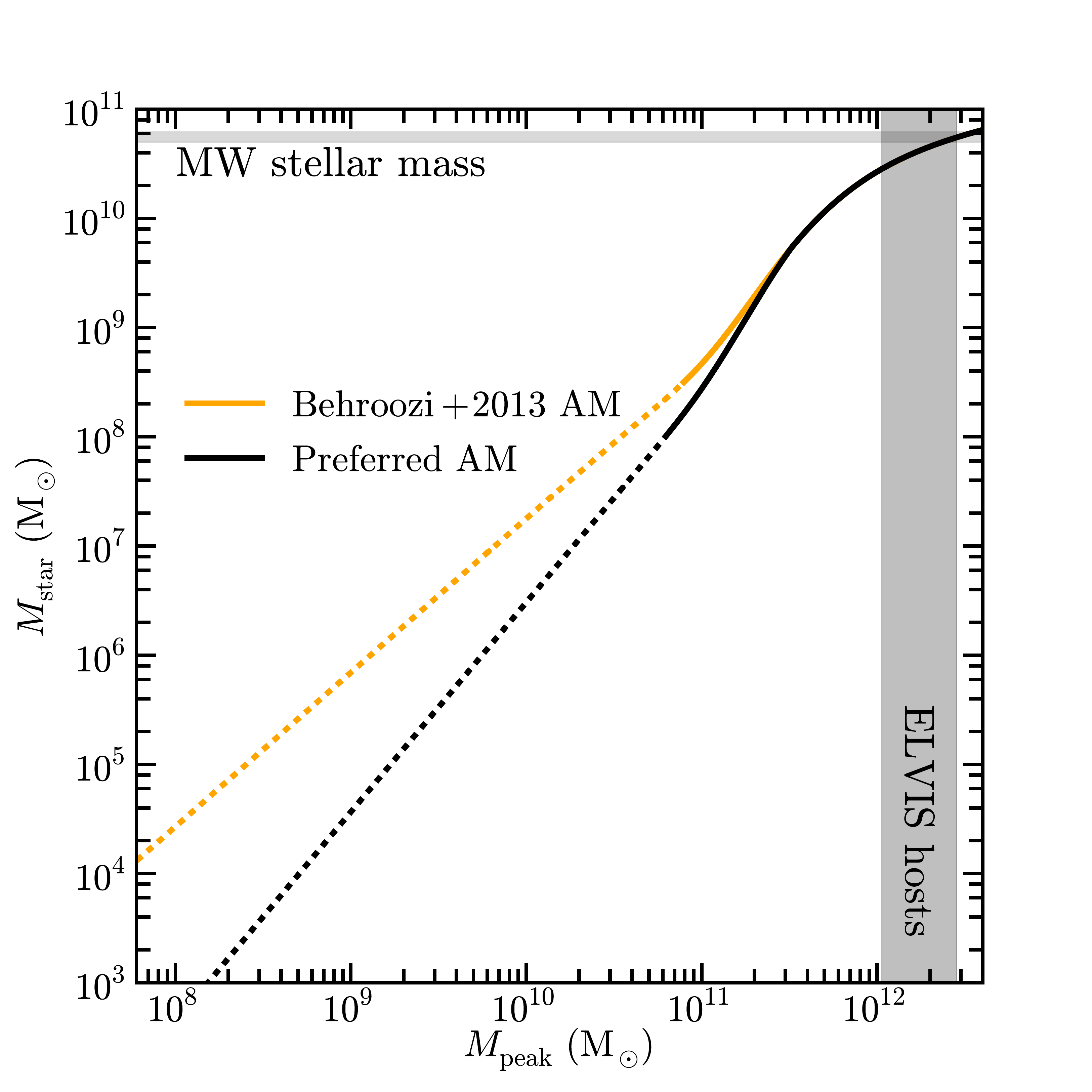}
  \caption{The abundance matching relation between stellar mass and halo mass
    from \citet[][orange line]{BehrooziAM}, extrapolated to low halo masses,
    compared to a modified relation (black) motivated by the updated stellar
    mass function of \citet{Baldry2012}.  As shown in
    Figure~\ref{fig:Mstar_300kpc}, this modified relation does a better
    job of reproducing faint ($\mstar \sim 10^6 - 10^8 \msun$) galaxies in the
    Local Group.  The two lines are solid over the range where the input stellar
    mass functions are complete and become dashed in the regime associated with
    pure extrapolation.  For reference, the horizontal grey band shows the
    stellar mass of the MW from \citet{Bovy2013}.  The virial masses of
    our ELVIS hosts span the vertical grey band.  Note that while our halo
    virial masses are consistent with dynamic estimates of MW and M31 virial
    masses, they are at the low-mass end of AM expectations for
    a system with the stellar mass of the MW.}
  \label{fig:AM}
\end{figure}

Figure~\ref{fig:AM} shows the $z=0$ AM relation published by \citet{BehrooziAM} as the 
orange line.  The plotted relation becomes dashed at $\mstar < 10^{8.5}\,\msun$, reflecting
the approximate completeness limit of the SDSS-derived stellar mass function of \citet{Baldry2008}, 
on which the \citet{BehrooziAM} relation was based.  The black line shows a modified version
of the \citet{BehrooziAM} relation, motivated by the updated stellar mass
function of \citet{Baldry2012}, who found flatter faint-end slope ($a_* = -1.47$
versus $-1.6$ in Baldry et al. 2008) using the Galaxy And Mass Assembly (GAMA)
survey \citep{GAMADR1}, which probes $\sim 2$ mag deeper than SDSS,
albeit over a smaller area of sky.  In this modified relation we have simply
altered the asymptotic slope $\alpha$ to be 1.92 in equation 3 of
\citet{BehrooziAM}, such that at small masses $\mstar \propto \mpeak^{1.92}$.
This is based on the expectation that $\alpha = (1+a_{\rm dm})/(1+a_*)$ and
assuming an asymptotic halo mass function slope of $a_{\rm dm} = -1.9$
\citep[e.g.][]{Jenkins2001}.  As we show below, this modified relation does a
better job in reproducing dwarf galaxy counts in the LG than the
original \citet{BehrooziAM} formulation.  Our preferred relation is
well described by a power law for $\mstar < 10^8 \msun$:
\begin{equation}
\mstar(\mpeak)= 3 \times 10^6 \, \msun \left(\frac{\mpeak}{10^{10} \msun}\right)^{1.92} \, .
\label{eqn:AMrelation}
\end{equation}
In the mass range of interest, this modified $\mstar-\mpeak$ relation 
is more similar to the AM prescription presented in \citet{Moster2013}.  
This relation is valid only at $z = 0$; our technique does not allow for a constraint at higher redshifts.

\begin{figure}		
  \centering
  \includegraphics[width=0.49\textwidth]{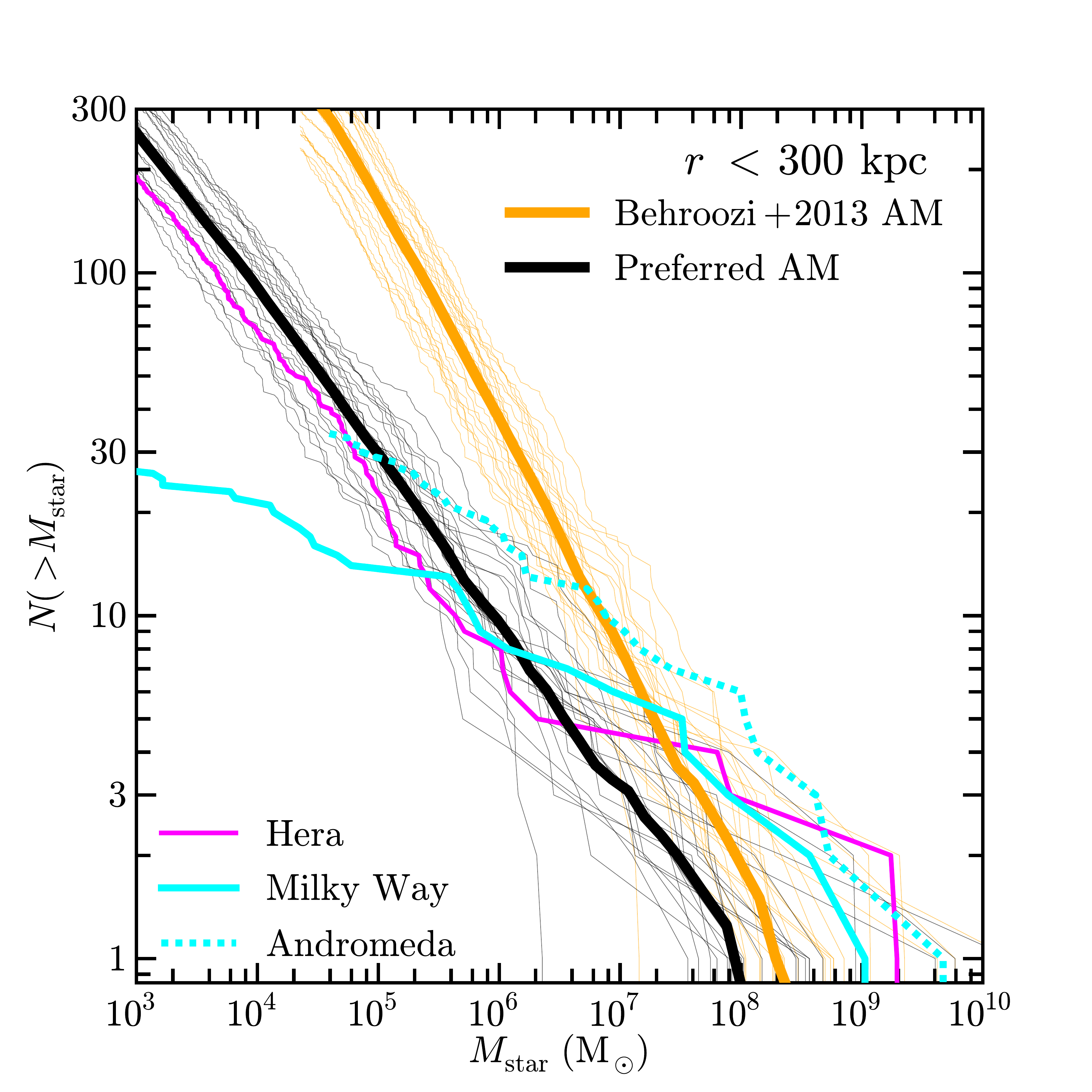}
  \caption{A comparison of observed stellar mass functions within 300~kpc of the
    MW (cyan) and M31 (dashed cyan) with predictions from the ELVIS subhalo catalogs and
    extrapolated AM relations.  The orange lines use the AM
    prescription of \cite{BehrooziAM}, which adopts a faint-end slope of the
    luminosity function of $-1.6$ \citep{Baldry2008}, while the black curves
    modify the Behroozi relation by assuming a slightly shallower faint-end
    slope of the luminosity function of $-1.47$ \citep{Baldry2012}.  The
    standard \cite{BehrooziAM} relation over-predicts the LG data significantly
    at $\mstar = 5 \times 10^5 \msun$, a regime where the census of satellites is
    believed to be complete.  The modified Behroozi relation (given in the text)
    is a better match to the observed counts.}
  \label{fig:Mstar_300kpc}
\end{figure}

Figure~\ref{fig:Mstar_300kpc} shows the stellar mass functions of galaxies
within 300 kpc of either the Milky Way (cyan) or M31 (dashed cyan) compared to the
predicted stellar mass functions for our ELVIS pairs based on the two AM
relations shown in Figure \ref{fig:AM}.  For the galaxy stellar mass functions,
we use the masses from \citet{Woo2008}, where available, and the luminosity data
cataloged in \citet[][]{McConnachie2012}, assuming $\mstar/L = 2$, otherwise.
The lines become dashed where incompleteness may become an issue \citep[see,
e.g.][]{Koposov2008,Tollerud2008,Richardson2011,Yniguez2013}.

The orange lines in Figure~\ref{fig:Mstar_300kpc} show the stellar mass
functions for each of the 24 paired ELVIS hosts derived from the
$\mstar(\mpeak)$ relation of \citet{BehrooziAM}.  The average relation is shown
by the thick line.  For this exercise we have applied the $z=0$ relation to all
subhaloes using their $\mpeak$ masses, which follows the prescription of
\citet{BehrooziAM}.  As can be seen, the standard \citet{BehrooziAM} relation
gives a stellar mass function that is too steep, over-predicting the count of
galaxies smaller than $\mstar \simeq 10^7 \msun$ significantly.  Our modified
relation (applied to ELVIS haloes in black) does a better job by assigning less
stellar mass to smaller haloes.  For this reason we will adopt this preferred AM
relation in all relevant figures to follow.  In magenta, we highlight the
satellites of the host Hera, which happens to be a particularly good match to 
the data (at least in the regime where it is likely complete) in this and several 
figures that follow.  Based on our preferred AM relation, we predict $\sim 200-300$ 
galaxies with $\mstar \geq 10^3\msun$ within 300~kpc of the MW / M31.  

We note that both AM prescriptions under-predict the satellite stellar mass function 
for the MW / M31 at $\mstar\geq10^8\msun$ when considering the average satellite mass 
function. At these relatively high masses, however, the halo-to-halo scatter is large 
and the well-established rarity of LMC-like objects 
\citep{Boylan-Kolchin2010,Busha2011,Tollerud2011} biases the mean result relative to 
observations of the LG. The stellar mass functions around individual hosts 
with large subhaloes, e.g. Hera in magenta, match observations well over four decades 
in stellar mass after applying the preferred AM relation.

\begin{figure}		
  \centering
  \includegraphics[width=0.49\textwidth]{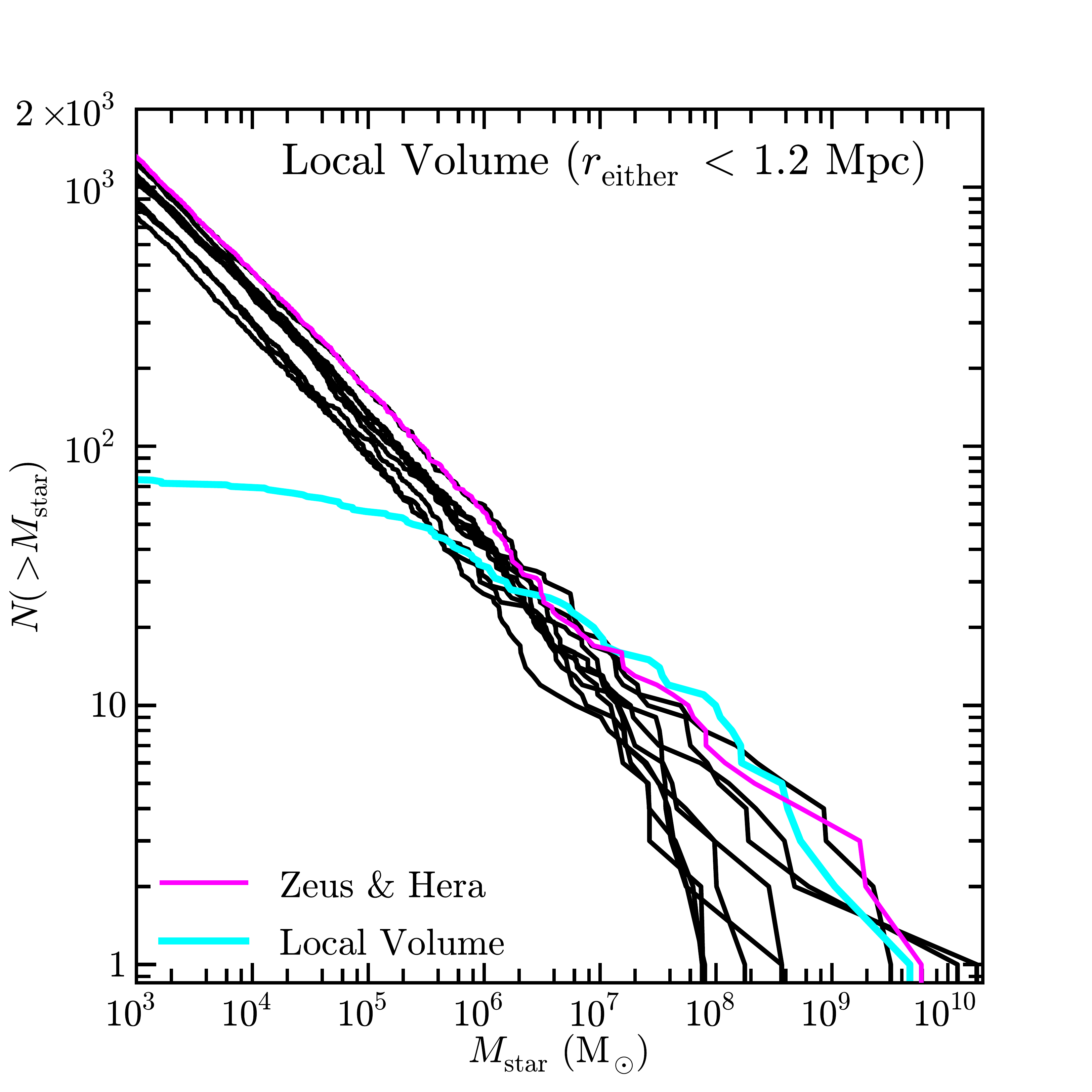}
  \caption{Cumulative stellar mass functions around paired hosts within the
    Local Volume using the preferred AM relation discussed in the text; not
    shown are those systems that include a third massive halo nearby (Siegfried
    \& Roy and Serena \& Venus).  The pair Zeus \& Hera are highlighted in
    magenta.  The current count of galaxies within the same volume around the MW
    and M31 is shown in cyan \citep{McConnachie2012}, which flattens at small
    mass, likely because of incompleteness.  We predict $\sim 1000$ galaxies having
    $\mstar \geq10^3\msun$ within this volume, compared to the $\sim70$
    currently known.}
  \label{fig:MstarFuncs}
\end{figure}

Figure~\ref{fig:MstarFuncs} presents stellar mass functions for simulated Local
Volumes (unions of 1.2 Mpc spheres around either host) using our preferred AM
relation. There is one line for each pair of haloes in the ELVIS sample,
excluding the two cases that contain a third large halo nearby (detailed in
\S~\ref{subsec:properties}).  Our AM-based prediction agrees reasonably well with the
data for $\mstar \gtrsim 5 \times 10^6\msun$, but rises much more steeply
towards lower masses, in the regime where the current census is almost certainly
incomplete.  We highlight the pair Zeus \& Hera in magenta.  This pair has an
$\mstar$ function that happens to be very similar to that of the LG. We see that
if the AM relation is extrapolated down to $\mstar \sim 10^3\msun$ we expect
$\sim 1000$ galaxies within the Local Volume (compared to the $\sim 70$ systems
currently known).  Future surveys like those performed with LSST \citep{LSSTWP}
will help test such extrapolations, exploring the relationship between halo mass
and galaxy mass at the very threshold luminosities of galaxy formation.

\subsection{HI Mass Functions}
\label{ssec:HImassfuncs}

While future resolved-star surveys promise to discover faint optical galaxies
throughout the Local Volume, HI surveys offer a complementary approach for the
discovery of dwarfs in the near-field \citep{Blitz1999, Blitz2000,
  Sternberg2002, Adams2013, Faerman2013}.  While the faintest dwarfs within
$\sim 300$ kpc of either the MW or M31 are gas-poor dSphs, gas-rich dwarfs are
the norm beyond the virial regions of either giant
\citep{Grcevich2009,McConnachie2012}.  Leo T, at distance of $\sim 400$ kpc from
the MW, is an example of a very faint system that is gas-rich \citep[$\mstar
\simeq M_{HI} \simeq 10^5 \msun$;][]{Ryan-Weber2008} and apparently falling in
to the MW virial radius for the first time \citep{Rocha2012}.  Similar, though
possibly even less luminous, objects may fill the Local Volume, and if so, could
be detected in blind searches for neutral hydrogen.  Recently, for example, the
gas-rich galaxy Leo P ($M_{HI} \simeq 3 \, \mstar \simeq 10^6 \msun$) was
discovered at a distance of $\sim 1.5-2$ Mpc using HI observations
\citep{Giovanelli2013,Rhode2013}.

Here we use the ELVIS suite to provide some general expectations for the HI mass
function in the Local Volume.  Building off of the results presented in
\S~\ref{ssec:mstarfuncs}, we use our preferred AM relation coupled with an
empirically-derived $\mstar$-$M_{\rm HI}$ relation to assign HI masses to haloes
in our simulated Local Volumes.  Specifically, we fit a power-law relation to
the gas-rich dwarfs in the LG from \citet{McConnachie2012}, ensuring
that the gas-fraction relation matches that found by \citet{Huang2012} at higher masses:
\begin{equation}
M_{\rm HI} = 7.7 \times 10^{4} \msun \, \left(\frac{\mstar}{10^5\msun}\right)^{1.2} \, .
\label{eqn:MstarMHI}
\end{equation}
Of course, this simple assumption of a one-to-one relation between stellar mass
and HI mass is highly idealized.  In reality, the gas-to-stellar-mass relation
shows a considerable amount of scatter \citep{Kannappan2004,
  McGaugh2005,Stewart2009,Huang2012, Huang2012dwarfs, Kannappan2013}, and this
is especially true for the faintest systems in the LG \citep[as
summarized in][]{McConnachie2012}.  A more realistic investigation of the HI
content of LG galaxies is reserved for future work.

We further assume that any halo that has been within the virial radius of a
giant has had all of its HI gas removed.  This presupposes that a process such
as ram pressure stripping removes the gas from satellites upon infall and is
motivated by observations demonstrating that the vast majority of Local Group 
satellites have negligible neutral gas content \citep{Grcevich2009}.
The small number of gas-poor dwarfs that lie beyond the virial radii of either
M31 or the MW (i.e. Cetus and Tucana) may very well be explained as backslash
haloes \citep[see][]{Sales2007,Teyssier2012}.  Of course, some of the largest satellite
galaxies in the LG (e.g. the LMC and NGC 205) are clearly able to retain HI for
a non-negligible period of time after infall.  This would suggest that our
assumptions will lead to some under-counting of HI-rich galaxies, primarily at
the highest masses.  Some never-accreted haloes, however, may have lost their
gas via interactions with other field objects or with the cosmic web \citep{Llambay2013},
which may lead to some over-counting at small masses.

\begin{figure}		
  \centering
  \includegraphics[width=0.49\textwidth]{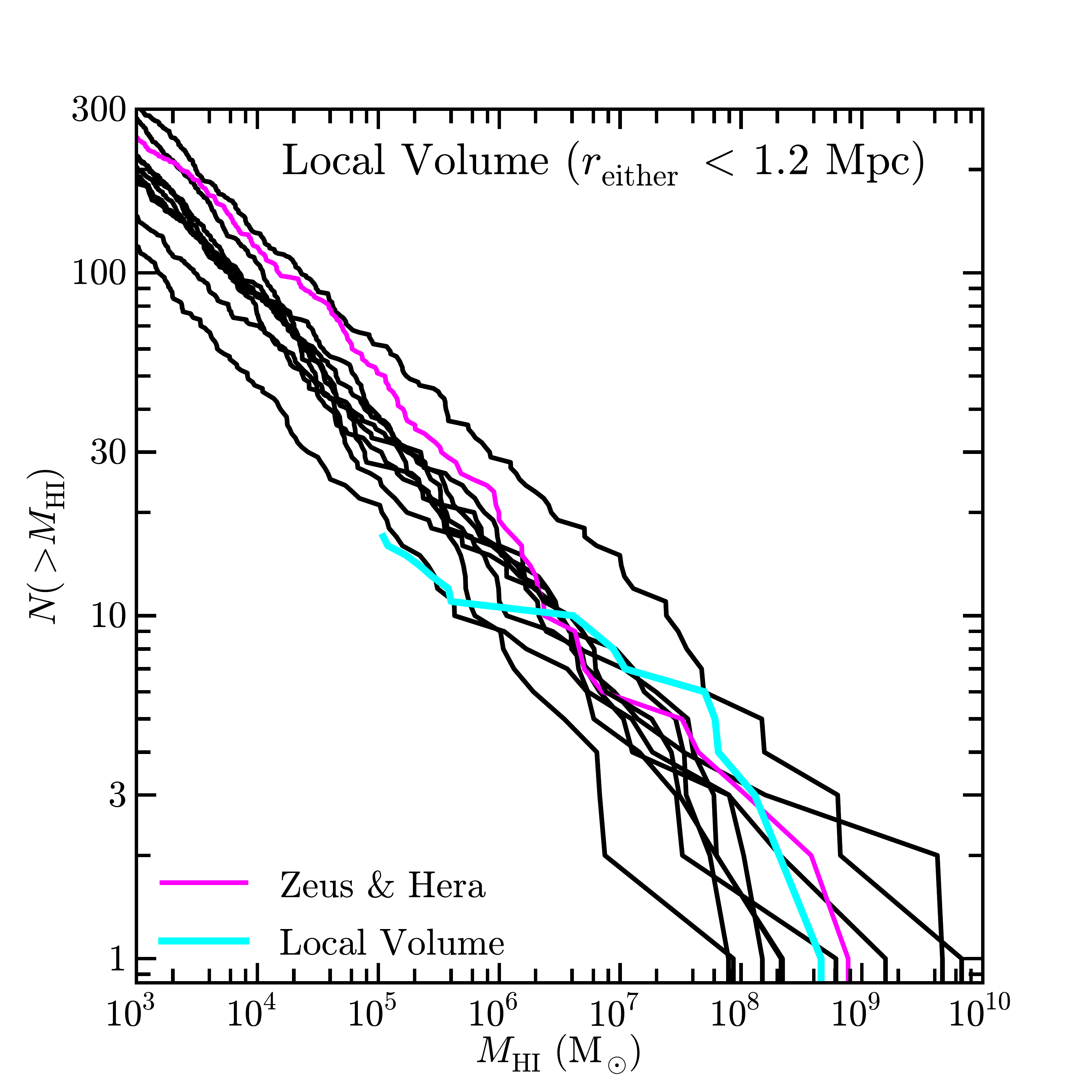}
  \caption{The HI mass functions within our simulated Local Volumes, excluding
    the systems with a third large host nearby.  We assign gas masses via
    Equation~\ref{eqn:MstarMHI}, assuming that any haloes that have passed within
    the virial radius of either giant since $z=5$ have been stripped of all gas.
    The local HI mass function is consistent for $M_{\rm HI} \gtrsim 5 \times
    10^6\msun$; below this value, incompleteness likely sets in.  We expect
    perhaps $\sim 50$ undiscovered galaxies with $M_{\rm HI} \geq 10^5\msun$
    within 1.2 Mpc of either host.}
\label{fig:Mgasfunc}
\end{figure}

The predicted HI mass functions within our simulated Local Volumes are plotted in 
Figure~\ref{fig:Mgasfunc}.  The two systems with large interlopers have
again been removed, and the line indicating Zeus \& Hera is again plotted 
in magenta.  The local HI mass function agrees well with predictions from ELVIS for 
$M_{\rm HI} \gtrsim 5 \times 10^6\msun$, at which point the local data break sharply, 
likely indicating incompleteness.  We estimate that there are as many as $\sim 50$ ($\sim 300$) unidentified 
galaxies with $M_{\rm HI} \gtrsim 10^5 \msun$ ($10^3\msun$) within 1.2~Mpc of the MW or M31.

\subsection{Compact High Velocity Clouds as Minihalos}
\label{subsec:alfalfa}
\begin{figure} 		
\centering
	\includegraphics[width=0.49\textwidth]{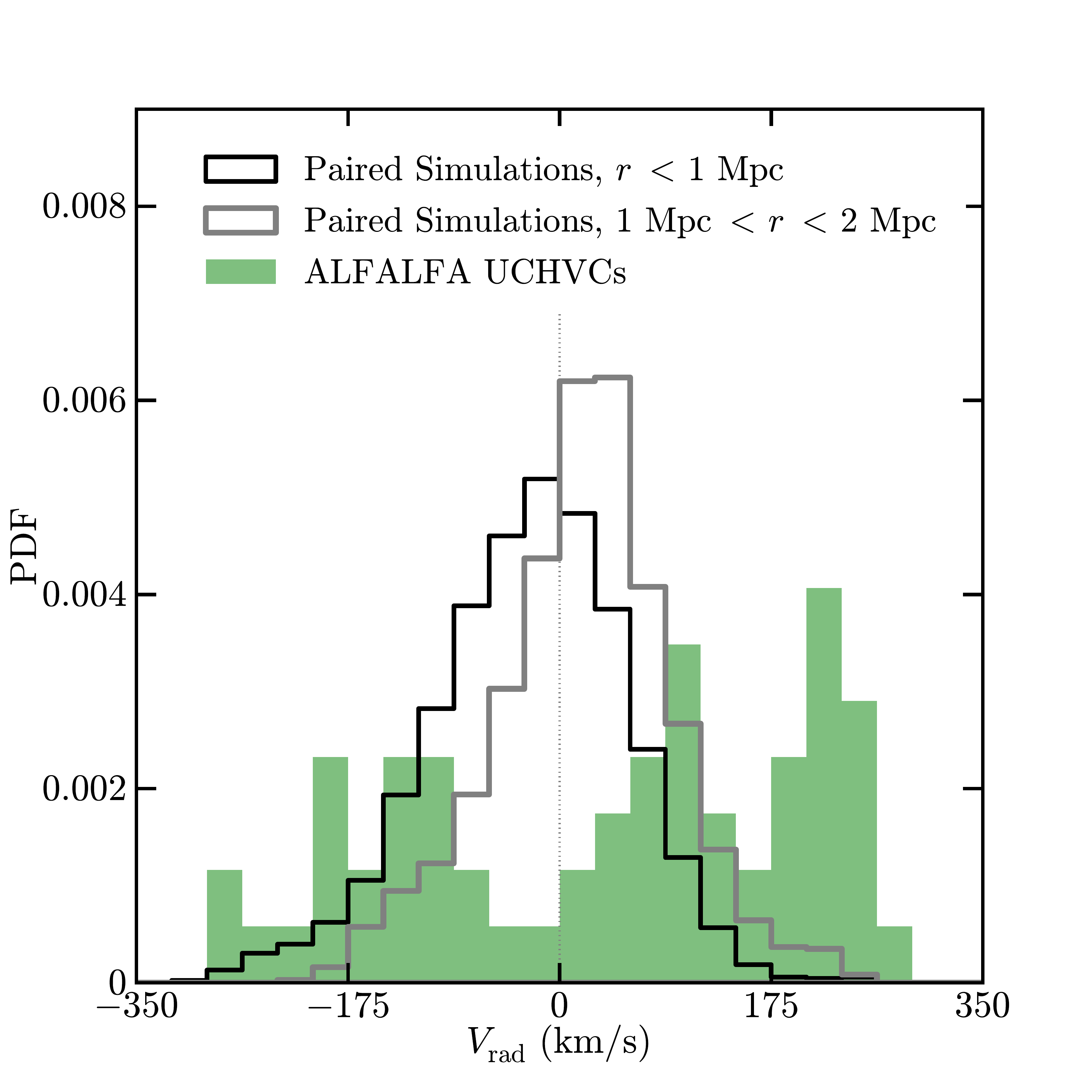}
        \caption{The black (grey) lines show the normalized radial velocity
          distribution of all predicted galaxies with $M_{\rm HI} > 10^{5}
          \msun$ within $1~\mpc$ ($2~\mpc$) of each host.  The shaded green histogram
          shows the radial velocities of the UCHVC halo candidates from
          \citet{Adams2013}.  While a selection bias limits the abundance of
          UCHVCs with $V_{\rm rad}\sim0~\kms$, the differences at the high radial
          velocity tail is illuminating.  Specifically, UCHVCs with $V_{\rm rad}
          > 175 \, \kms$ are highly unlikely to be associated with small haloes
          in the Local Volume according to our predictions.  Systems with lower
          radial velocities are likely better candidates for follow up.}
\label{fig:Vhistgas}
\end{figure}

It is possible that some of these gas-rich objects have already been detected.
Recently, \citet*{Adams2013} presented a catalogue of ultra-compact high velocity
clouds (UCVHCs) extracted from the Arecibo Legacy Fast ALFA 
\citep[ALFALFA;][]{ALFALFA,ALFALFA40} survey and discussed the possibility that 
some of these objects may be dwarf galaxies \citep[see also][]{Blitz1999,Faerman2013}.  
\citet{Adams2013} present 53 candidates, with HI properties that correspond to sizes 
of $\sim 3$ kpc and masses of $M_{\rm HI} \simeq 10^5 - 10^{6} \msun$ if they reside 
at $\sim 1$ Mpc distances.  These characteristics are suggestively similar to those of 
known LG galaxies like Leo T.  The ELVIS suite can be used to test whether these
UCHVCs have properties (radial velocities and overall counts) that are
consistent with those expected in \lcdm\ for small haloes in the Local Volume.

From Figure~\ref{fig:Mgasfunc}, we can immediately see that it is unlikely that
{\em all} of the \citet{Adams2013} candidates are associated with small dark
matter haloes in the Local Volume.  We expect fewer than $100$ undiscovered
objects {\em over the whole sky} with $M_{\rm HI} \gtrsim 10^5 \msun$ within 
1.2~Mpc of either host, while the ALFALFA sample has 53 candidates over only
$\sim10\%$ of the sky.  Nevertheless, it would not be surprising if some of the
identified candidates are indeed associated with dark-matter-dominated dwarfs.

The observed radial velocities of these clouds may provide clues for selecting
the best candidates for follow-up. Figure~\ref{fig:Vhistgas} shows the
normalized stacked radial velocity distribution of $M_{\rm HI} > 10^5 \msun$
haloes that sit between $\rvir$ and 1~Mpc (black curve), and between 1~and~2~Mpc
(grey curve) of our ELVIS pairs, measured from the centre of each host.  We again
exclude those objects with a third nearby giant from the black curve, and include
only those objects with $R_{\rm res} > 1.75~\mpc$ (Zeus, Charybdis, Romulus, and Kek) in
the grey curve, so as to minimize the effects of contamination from low resolution 
particles.  The shaded green histogram shows the radial velocity distribution of candidate 
mini-haloes from the \citet{Adams2013} UCHVC sample.  It is important to recognize 
that the UCHVC sample is biased to avoid the region near $V_{\rm rad} \approx 0\,\kms$ 
by construction.  Nevertheless, the high-velocity tail of distribution shows some
interesting differences compared to the predicted distribution.

The most important distinction between the simulated haloes and the candidate
objects is that there is a substantial population of UCHVCs with $175\,\kms \la V_{\rm
  rad} \la 350\,\kms$.  There are very few haloes predicted with such high
recessional velocities within 1~Mpc, and only slightly more out to 2~Mpc. We
conclude that the sub-population of UCHVCs with these high velocities is
unlikely to be associated with dark matter haloes unless they are substantially
more distant than 2 Mpc (in which case their total gas mass would become very
large and thus the expected count would drop considerably).  Based on these
results, we suggest that targeted follow-up searches for nearby mini-haloes may
want to focus on UCHVC candidates with $V_{\rm rad} \la 150 \,\kms$.

\begin{figure*} 	
  \centering
  \includegraphics[width=0.98\textwidth]{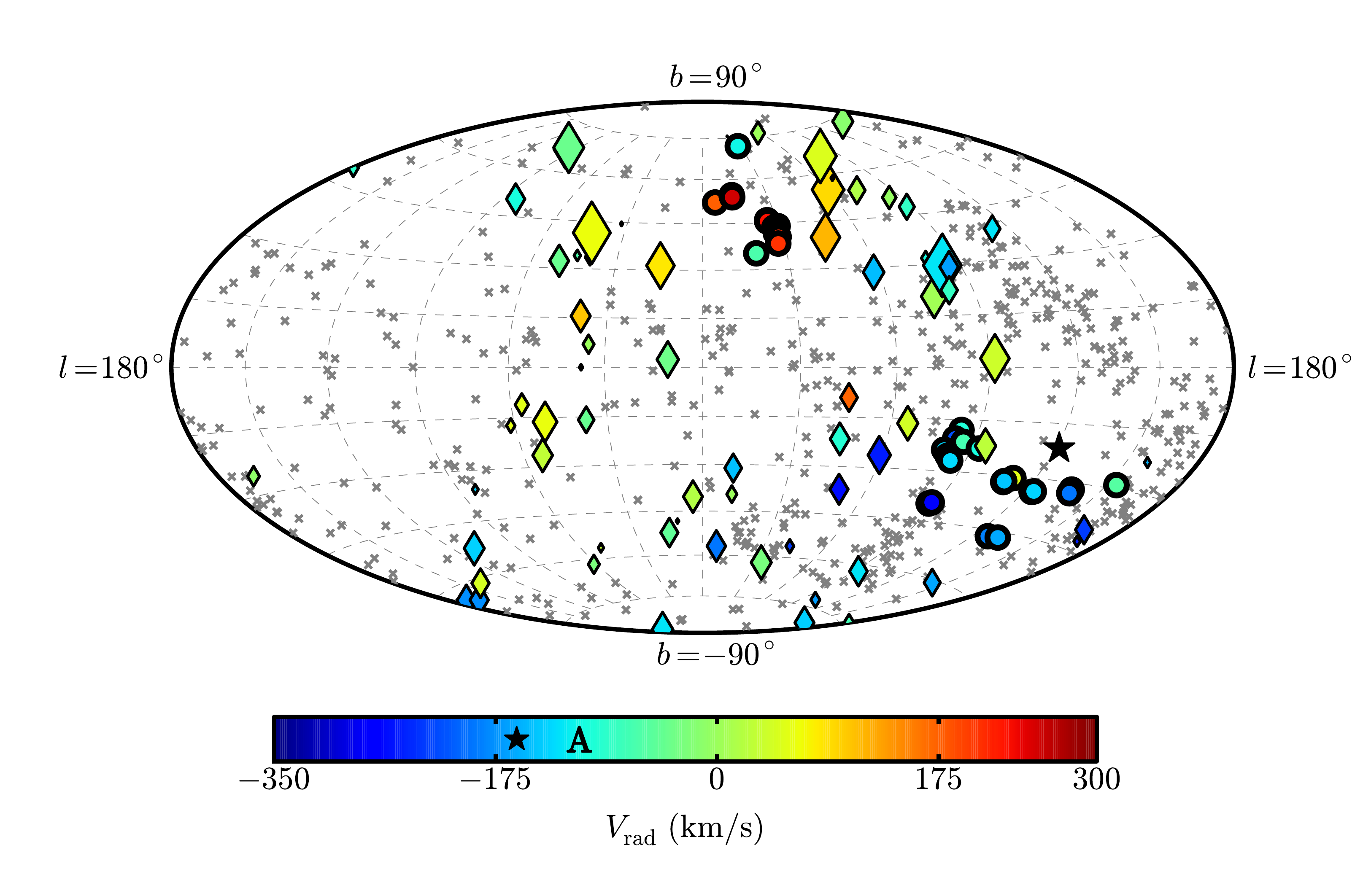} 
  \caption{ A Hammer projection of the haloes within 1 Mpc of Hera in mock
    Galactic coordinates, highlighting the haloes we expect to be gas rich with
    diamonds and marking backsplash haloes with crosses; no subhaloes of either
    giant are plotted.  The simulation is rotated such that Zeus and M31 lie in
    the same position on the sky; this point is marked with a star.  The size of
    the diamonds is proportional to our modeled gas mass values and distances as
    $\log(M_{\rm HI}/r^2)$. The UCHVC minihalo candidates from \citet{Adams2013}
    are plotted as circles with thick outlines.  These and the gas rich objects
    are coloured by their radial velocities according to the colour bar; the
    approach velocities of Zeus and Andromeda are indicated on the colour bar
    by the star and the A, respectively. The velocities of the fastest
    (outflowing) UCHVCs in the north are clearly outliers compared to the
    expected velocities of haloes in this region and therefore may be poor
    candidates for follow-up to discover dwarf galaxies.  The infalling systems
    in the south are more in line with our kinematic expectations for
    mini-haloes.}
  \label{fig:Jayaitoff}
\end{figure*}

We also compare the on-the-sky positions of the possible minihalos around the
Milky Way to those of the gas rich objects near Hera, the host that we have
highlighted throughout this work, in a Hammer projection in
Figure~\ref{fig:Jayaitoff}.  The diamonds indicate the predicted galaxies around
Hera and the circles denote the minihalo candidates from ALFALFA; both are
coloured by their relative radial velocities according to the colour bar.  We have
oriented the coordinate system such that Hera's partner halo Zeus sits at the
$(l,b)$ of M31 (indicated by the star).  The grey crosses are backsplash haloes.
There is a clear clustering of backsplash objects near Zeus and a corresponding
dearth of gas-rich haloes.  Suggestively, the receding ALFALFA objects, which
seem most inconsistent with the velocity distributions in ELVIS, are located
near one another.  We do note, however, that the gas clouds identified by
ALFALFA may instead be more distant objects that are perhaps still a part of the
Hubble Flow.  We find that most objects more than 1.5~Mpc from the centre of
each host are receding.

\subsection{The Local $\mathbf{r}-\mathbf{V_{\rm r}}$ Relation}
\label{subsec:hubblerelation}

The velocity field within the Local Volume contains a wealth of information on
the assembly history and mass of the Local Group \citep{Kahn1959,
Karachentsev2002, Karachentsev2005,Peirani2006, Teyssier2012,Marel2012}.  The
ELVIS simulations supply a potentially valuable basis for interpreting these
data, and we intend to utilize them for this purpose in future work.  Here we
briefly examine the local velocity-distance relation in one of our simulations
in order to demonstrate broad agreement with data and illustrate the potential
for a more in-depth interpretive analysis.

\begin{figure*} 	
  \centering
  \includegraphics[width=0.98\textwidth]{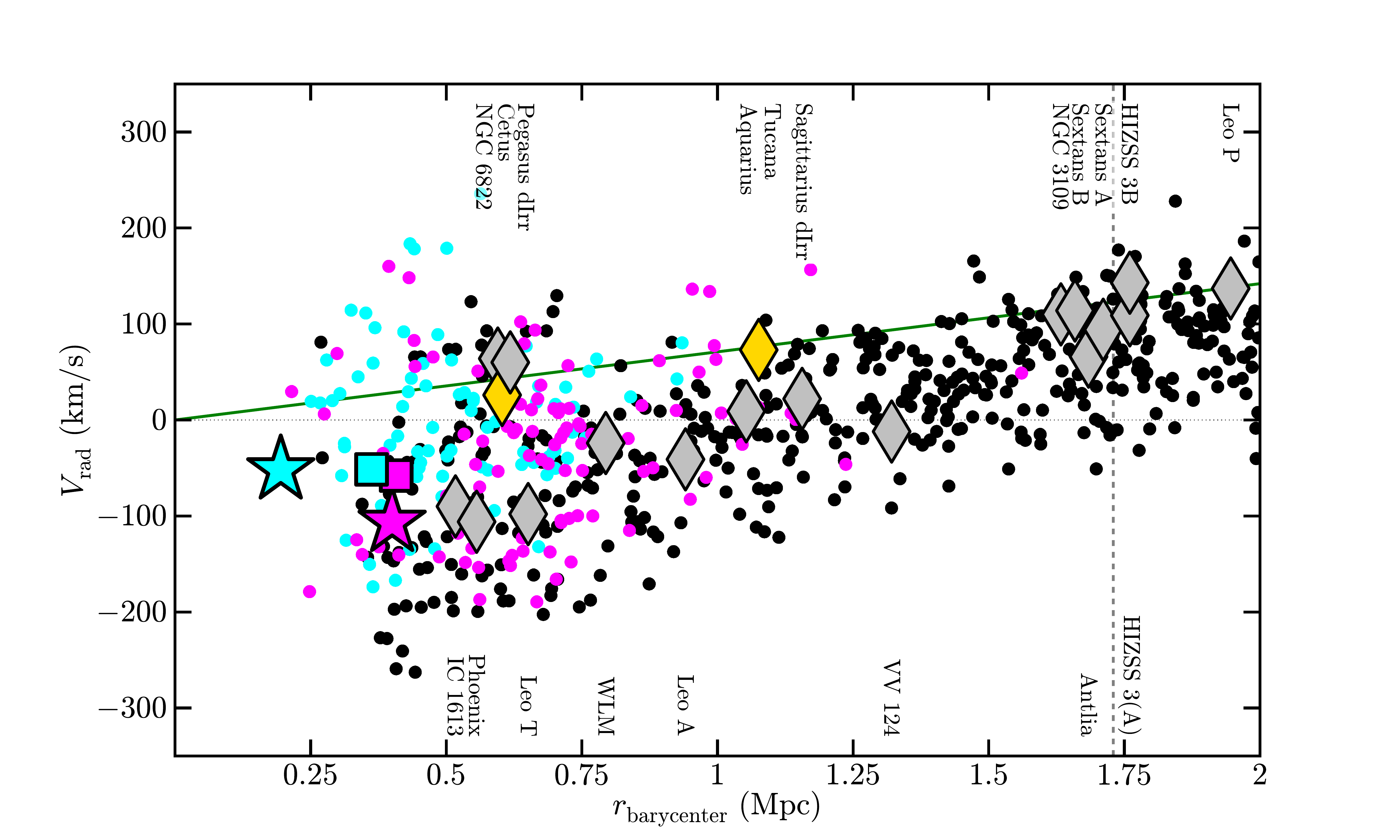} 
  \caption{Barycentric radial velocity versus barycentric distance for Local Group galaxies 
  compared to expectations from the Zeus \& Hera simulation.  The centers of Zeus (cyan) 
  and Hera (magenta) are indicated by large stars, while M31 (cyan) and the MW (magenta) are shown 
  by squares.  All haloes with $\mstar > 3\times10^3 \msun$ and beyond 300~kpc of 
  either Zeus or Hera are plotted as circles.  Large diamonds indicate known galaxies in 
  the Local Group beyond 300~kpc of either giant.  Backsplash haloes of Zeus are shown 
  as cyan while those that have interacted with Hera are plotted in magenta.  The black points 
  are haloes that have yet to be accreted by either host.  The grey diamonds mark LG 
  galaxies that have gas while the yellow diamonds correspond to Cetus and Tucana, 
  two gas-free dwarf that are good backslash candidates.  For reference, the green line 
  shows the asymptotic linear Hubble relation for our simulated cosmology.}  
  \label{fig:Hubble}
\end{figure*}

Figure \ref{fig:Hubble} shows the local relation between distance and radial velocity, 
centred on the Local-Group barycenter, along with data from the Zeus \& Hera simulation.  
MW and M31 are indicated as magenta and cyan squares, respectively, calculated from the
separation and radial velocity given in Table~\ref{tab:pairprops} and the masses in
Table~\ref{tab:solopairprops}.  Known Local Group galaxies that reside beyond
300 kpc of either giant are shown as large diamonds; the two highlighted in
yellow are the gas-free dwarfs Cetus and Tucana, which are backsplash
candidates.  The Leo P data point is calculated from Tables~\ref{tab:pairprops}
and~\ref{tab:solopairprops}, assuming that its distance from the MW is 1.75~Mpc
\citep{McQuinn2013}; the remainder of the observational data is taken from
\citet{McConnachie2012}.  For comparison, circles show all haloes within the Zeus
\& Hera simulation that are large enough, according to our preferred
AM relation, to have stellar masses exceeding 3000 $\msun$.
Halos within 300 kpc of either simulated giant are excluded, but galaxies that
have been within the virial radii of Zeus or Hera are coloured cyan and magenta,
respectively.

As expected from the previous discussion (e.g. Figure 8), backsplash haloes tend
to populate the outflowing, positive-velocity envelope of the relation.  The
gas-free dwarfs Cetus and Tucana are similarly consistent with inhabiting this
upper envelope.  More generally, the simulated relation is a reasonably good
match to the data shown.  The relative lack of known galaxies with $V_{\rm r}
\lesssim -100\,\kms$ is likely related to the barycentric velocities of the MW
and M31, which are moving $\sim50\,\kms$ slower than Zeus and Hera.

Finally, we note that the vertical dashed line near 1.75~Mpc in Figure
\ref{fig:Hubble} indicates the position of the first low-resolution
(contamination) particle in the simulation.  In principle, our predictions could
be compromised beyond this point, but based on larger (lower resolution)
simulation comparisons we find no evidence that contamination biases bulk
velocity predictions.

\section{Conclusions}
\label{sec:conclusions}
This work presents the ELVIS suite, a set of collisionless cosmological
simulations consisting of 12 Local Group-like pairs of MW/M31-size dark
matter haloes and 24 isolated analogs mass-matched to those in the
pairs.  Each simulation resolves mini-haloes down to $\mpeak = 6 \times 10^7
\msun$ within high-resolution, contamination-free volumes that span 2 to 5 Mpc
in size.

One of the goals of this work is to determine if the Milky Way and M31 are
expected to be biased in any way with respect to typical field haloes as a result
of their paired configuration.  We find no evidence that this is the case
(c.f. Figure~\ref{fig:MpRvir}).  Statistically, subhalo properties (counts and 
kinematics) and host halo properties (formation times and concentrations) are
indistinguishable between our paired and unpaired samples.  We provided 
analytic fits to subhalo mass functions in \S~\ref{subsec:abundance} (and for 
$\vmax$ functions in Appendix~B).  Apparently, as long as measures are 
restricted to the virial volumes, simulated field haloes provide an adequate 
comparison set for the MW and M31.

As might be expected, differences become more apparent between paired and
isolated samples when we explore measures beyond the virial volumes of either
hosts (Figures~\ref{fig:MpDenver}--\ref{fig:Vradhist}).  The Local Volume at 
$1.2$~Mpc distance around each paired host contains, on average, 80\% more haloes 
at fixed $\mpeak$ than the corresponding region surrounding each isolated host.  
Similarly, the kinematic properties of the mini-halo population around LG -like 
pairs show distinct differences from isolated MWs: the tangential 
velocity distributions for haloes around pairs are significantly hotter, and the 
radial velocity distributions are skewed towards more outflowing systems.  The 
tendency to see more outwardly moving haloes around paired hosts is likely related 
to another difference we see: an increase in the backsplash fraction.  We find 
evidence that paired haloes have an increased fraction of satellite systems that are 
now beyond the virial radius of either host, but that had previously been
inside (Figure~\ref{fig:fbacksplash}).  These backsplash objects are preferentially 
moving outward along more radial orbits at $z=0$ (Figure~\ref{fig:Vhist48}).

With these basic comparisons in place, we investigate our sample of LG-like pairs 
more closely, focusing on comparisons with data throughout the Local Volume.  
A summary of the resultant work is as follows:
\begin{itemize}
\item We find that the abundance matching relation presented by
  \citet{BehrooziAM} over-predicts the number of $M_\star \sim 5 \times 10^6
  \msun$ satellites within 300~kpc of the MW and M31 (Figure~\ref{fig:Mstar_300kpc}), 
  a regime where the satellite census is believed to be complete.
\item We present a modified Behroozi relation, motivated by the stellar mass
  function reported by \citet{Baldry2012} from GAMA data (Figure~\ref{fig:AM} 
  and Equation~\ref{eqn:AMrelation}), that reproduces the observed satellite 
  count down to $M_\star \sim 5 \times 10^5\msun$, a point where incompleteness 
  likely becomes an issue.  It also reproduces galaxy counts throughout the Local 
  Volume down to $M_\star \sim 5\times 10^6 \msun$, below which incompleteness is 
  almost certainly an issue (Figure~\ref{fig:MstarFuncs}).
\item By extrapolating our preferred AM relation to low halo
  masses, we find there should be $\sim300$ galaxies with $\mstar \geq
  10^3\msun$ within 300 kpc of the Milky Way and $\sim1000$ such galaxies within
  1.2~Mpc of either host.  LSST (along with ongoing surveys) will test this
  expectation.  If faint galaxies are not discovered in large numbers, it could
  point to a break in the stellar-mass to halo-mass relation at the low-mass end.
\item Using empirical relations between HI mass and stellar mass, we predict the
  number of gas-rich galaxies within the Local Volume (Figure~\ref{fig:Mgasfunc}).  
  The observed LG HI mass function agrees well with our expectations down to 
  $M_{\rm HI} \sim 10^7\msun$, below which the data may suffer from incompleteness.  
  We conclude that there may be approximately 50 undiscovered gas-rich haloes with 
  $M_{\rm HI} > 10^5 \msun$ within 1.2 Mpc of the Milky Way and M31.
\item We compare the properties of our modeled gas-rich haloes to the UCHVC
  mini-halo candidates presented by ALFALFA \citep[][Figures~\ref{fig:Vhistgas}~and~\ref{fig:Jayaitoff}]{Adams2013}.  
  While the characteristics of many of these clouds make them good candidates for gas-rich
  haloes, it is highly unlikely that more than $\sim 10 \%$ are true mini-haloes.
  In particular, positive radial velocities in excess of 175 $\kms$ are drastically
  inconsistent with our expectations for halo kinematics within $\sim 2$ Mpc of
  the MW.
\end{itemize}

Our results generally indicate that studies focusing on basic properties within
the virial volumes of the MW or M31 can be fairly compared to predictions from
more isolated field-halo simulations
\citep[e.g.][]{Diemand2008,Kuhlen2008,Aquarius}.
However, simulations investigating the volume
surrounding the Milky Way \textit{must} account for the overall environment that
it lives in -- specifically, the presence of the approaching M31 galaxy.
\vskip2cm

\noindent {\bf{Acknowledgments}} \\
Support for this work was provided by NASA through a \textit{Hubble Space
  Telescope} theory grant (program AR-12836) from the Space Telescope Science
Institute (STScI), which is operated by the Association of Universities for
Research in Astronomy (AURA), Inc., under NASA contract NAS5-26555.  This work
was also funded in part by NSF grants AST-1009999, AST-1009973, and NASA grant
NNX09AD09G.  M.B.-K. acknowledges support from the Southern California Center
for Galaxy Evolution, a multi-campus research program funded by the University
of California Office of Research.  J.S.B. was partially supported by the Miller
Institute for Basic Research in Science during a Visiting Miller Professorship
in the Department of Astronomy at the University of California Berkeley. 

The authors thank Frank van den Bosch, Michael Cooper, Manoj Kaplinghat, Evan
Kirby, Jose O\~{n}orbe, Julio Navarro, Annika Peter, and Risa Wechsler for
helpful comments, and Erik Tollerud for aid in creating the visualizations.  We
also thank Volker Springel for making \texttt{Gadget-2} publicly available and
for providing a version of \texttt{Gadget-3} for our use, Peter Behroozi for
making \texttt{Rockstar} and \texttt{consistent-trees} publicly available, and
Oliver Hahn for making \texttt{MUSIC} publicly available.  Finally, we
gratefully acknowledge the computational support of the NASA Advanced
Supercomputing Division and the NASA Center for Climate Simulation, upon whose
\textit{Pleiades} and \textit{Discover} systems these simulations were run, and
the \textit{Greenplanet} cluster at UCI, upon which much of the secondary
analysis was performed.

\section*{Appendix A: Numerical Convergence}
\begin{figure*}
  \centering
  \includegraphics[width=0.49\textwidth]{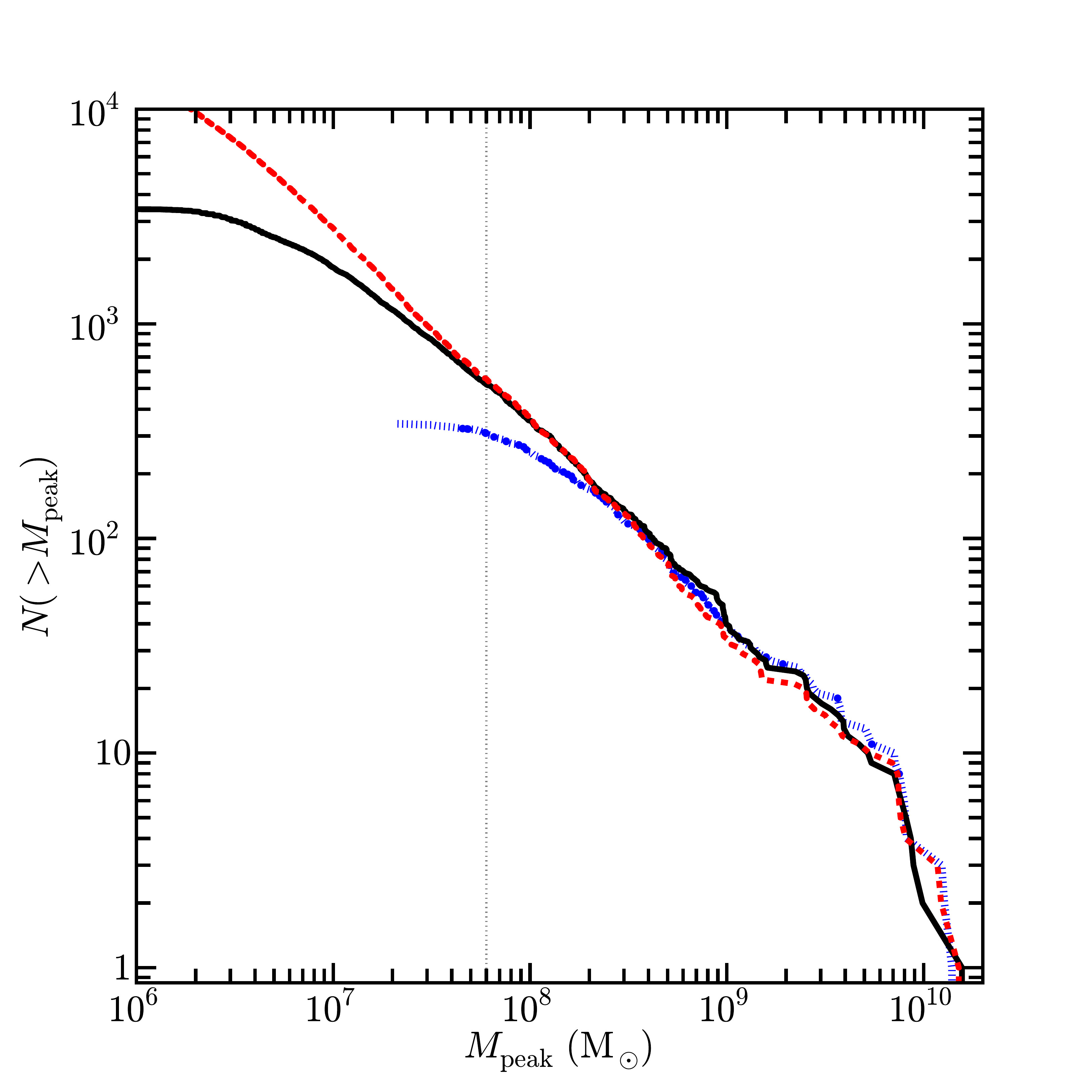}
  \centering
  \includegraphics[width=0.49\textwidth]{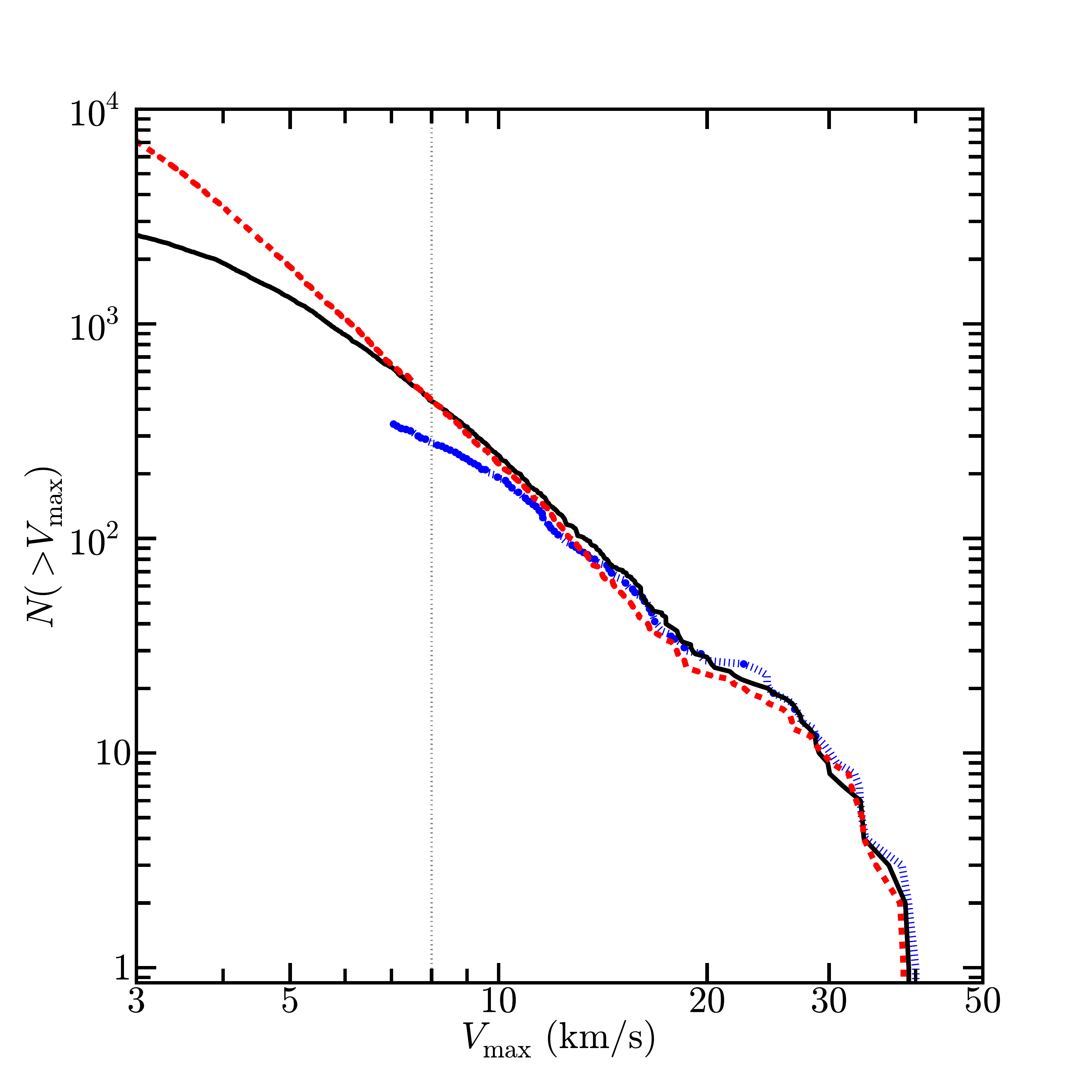}
\caption{Resolution test indicating the smallest haloes 
    \texttt{Rockstar} reliably identifies in the ELVIS simulations.  Here 
    we plot the $\mpeak$ (left) and $\vmax$ (right) functions for haloes 
    within $400$~kpc of the smallest of our isolated haloes, iKauket.  The black 
    line indicates the fiducial resolution; the red line is from the HiRes 
    simulation, and the blue line is from a lower resolution run, for 
    illustrative purposes.  The mass and circular velocity at which the lines 
    begin to systematically disagree, $\mpeak = 6\times 10^7\msun$ and 
    $\vmax=8\,\kms$, constitute our resolution limits for the fiducial 
    resolution.}
\label{fig:res_test}
\end{figure*}

In this Appendix, we compare the $\mpeak$ and $\vmax$ functions within
$400\,\kpc$ of iKauket at three different levels of numerical
resolution. Figure~\ref{fig:res_test} contains this comparison: results from the
HiRes simulation ($m_p=2.35\times 10^{4}\,\msun$, $\epsilon=70.4\,{\rm
  pc}$) are shown as a red dashed line, while results from the run at our
fiducial resolution ($m_p=1.89\times 10^{5}\,\msun$,
$\epsilon=141\,{\rm pc}$) are shown as a solid black line. For
comparison, the blue line shows a lower resolution run as well
($m_p=1.55\times10^6\msun$, $\epsilon=469\,{\rm pc}$).

The left panel plots the number of haloes identified by our pipeline with
$\mpeak$ greater than a given mass; on the right, we plot the current $\vmax$
function. By locating where our fiducial resolution begins to systematically
differ from the HiRes run, it is clear that haloes with $\vmax > 8$~km/s and
$\mpeak > 6\times 10^7\msun$ are reliably identified at the fiducial
resolution. These resolution limits are marked by dashed vertical lines
in the plots.

\section*{Appendix B:  $\mathbf{\vmax}$ Functions}
\begin{figure}
\centering
\includegraphics[width=0.49\textwidth]{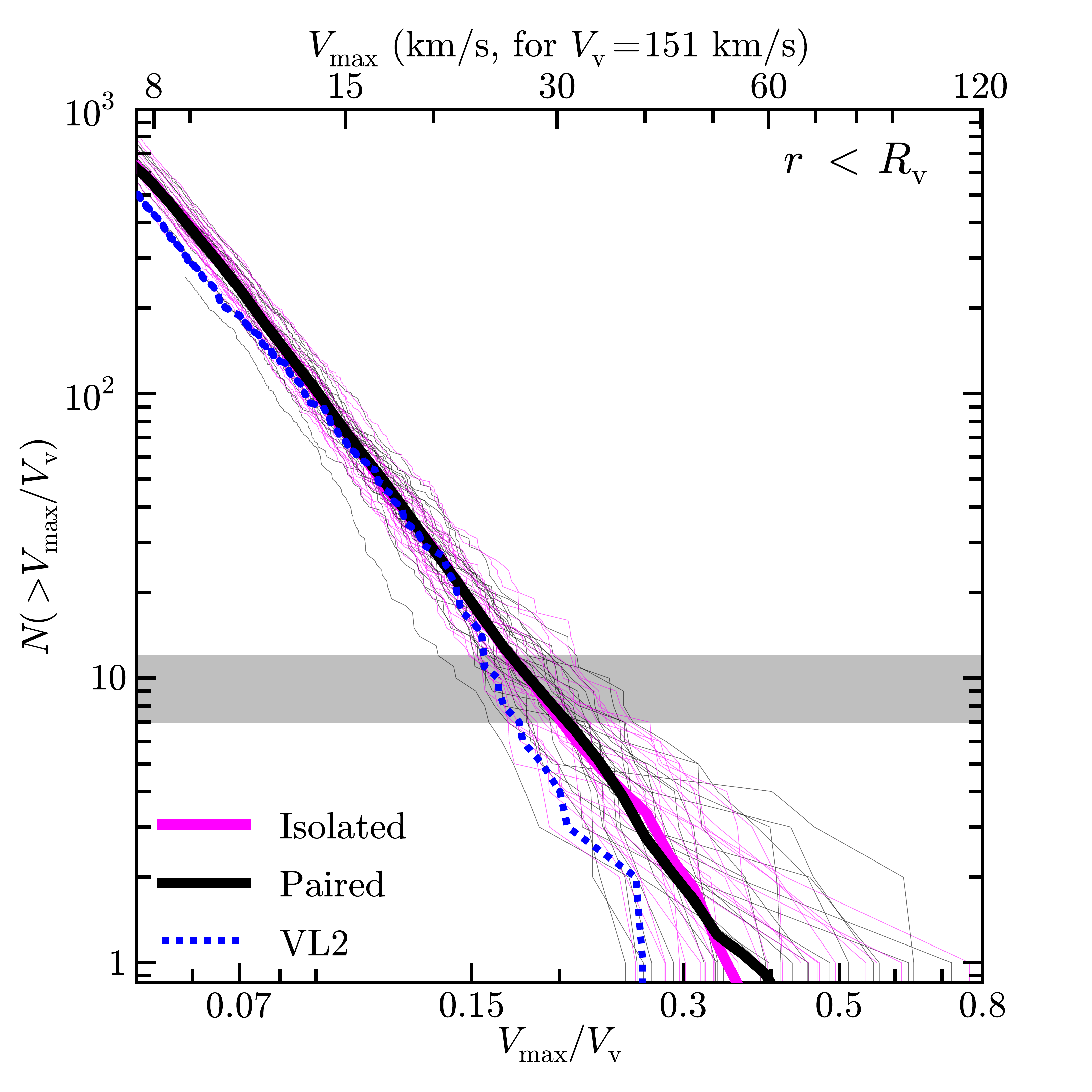}
\caption{The $\vmax$ function within $\rvir$ of each host, scaled by the virial
  velocity of that host, analogous to Figure~\ref{fig:MpRvir}.  As in that
  Figure, the two populations agree well within the virial radius and are both
  well fit at the low-mass end by a power law of slope -3.1, as given in the
  text.  The blue dashed line plots the $\vmax$ function within the virial radius
  of the high-resolution Via Lactea II halo \citep{VL2}, which agrees within the
  halo-to-halo scatter.}
\label{fig:vmaxrvir}
\end{figure}

\begin{figure}
\centering
	\includegraphics[width=0.49\textwidth]{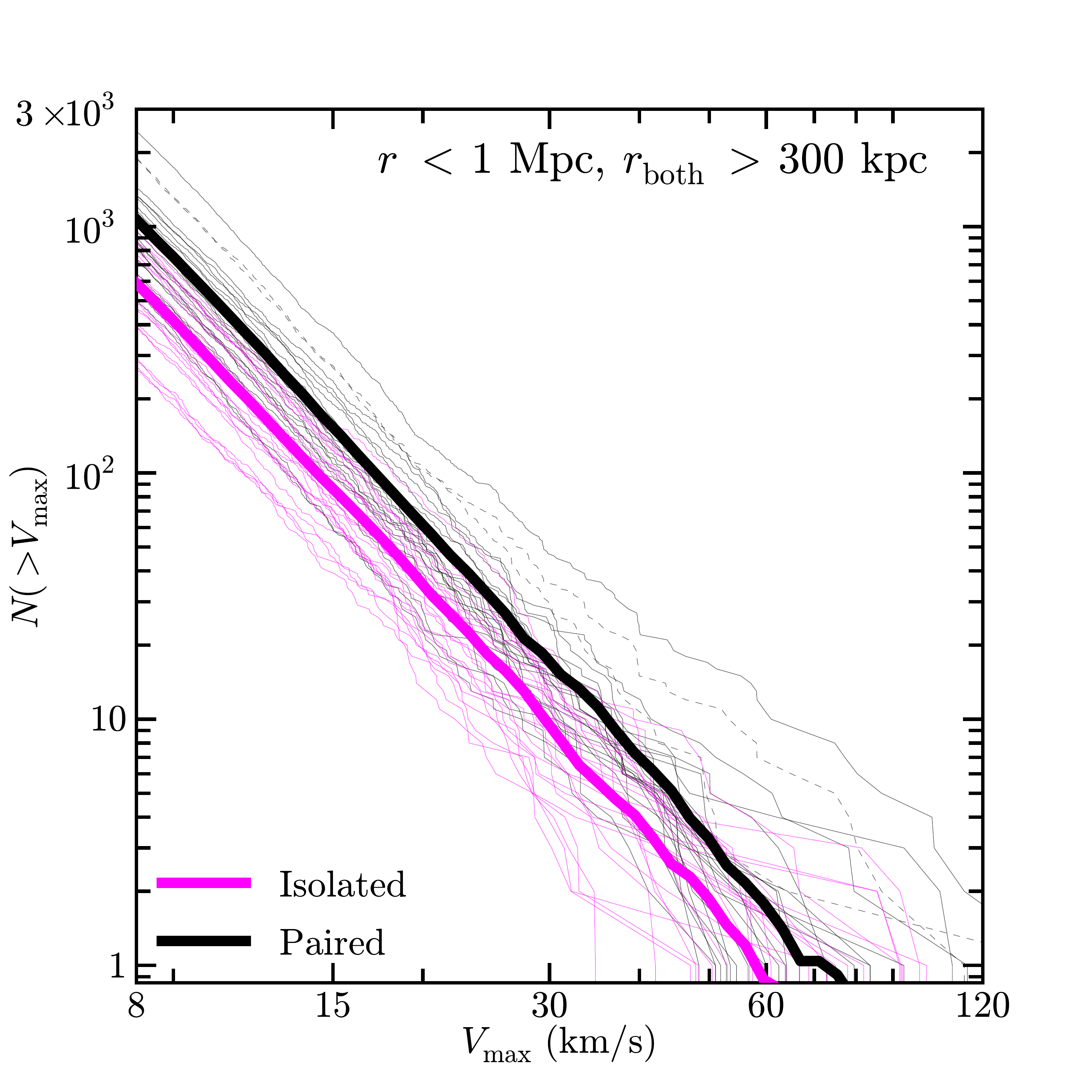}
        \caption{The $\vmax$ functions for objects in the Local Field (within
          1~Mpc of the host, but more than 300~pc from both giants).  The
          average relations are offset from one another, with the paired
          simulations having an amplitude that is 75\% higher.  The power law
          fits to the average relations are given in the text.}
\label{fig:vmaxdenver}
\end{figure}

\begin{figure}
\centering
	\includegraphics[width=0.49\textwidth]{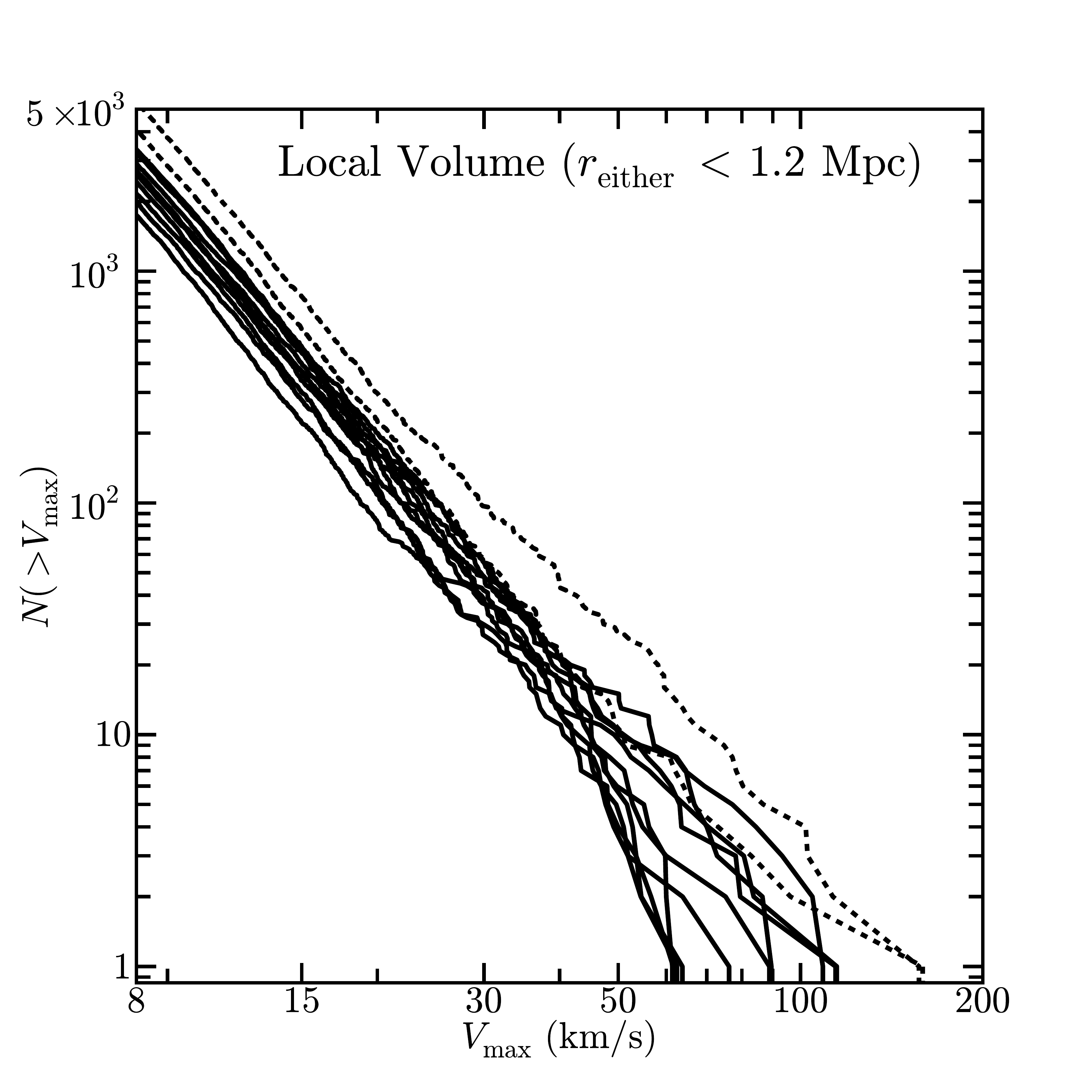}
        \caption{The $\vmax$ functions in the Local Volume (1.2~Mpc of either
          host), analogous to Figure~\ref{fig:MpMpc}.}
\label{fig:vmaxmpc}
\end{figure}

For most galaxies, it is more convenient to measure circular velocities or
velocity dispersions than virial mass.  Although we do show stellar mass
functions in the main body, our relation is not a mapping between $\mstar$
and $\vmax$; thus, we show $\vmax$ functions for direct comparison with
such observations here.  As with the $\mpeak$ functions, counts as a function
of $\vmax$ agree well within $\rvir$ (Figure~\ref{fig:vmaxrvir}), and are both 
well fit by a power law at the low mass end:
\begin{equation*}
N_{\rm v}(>V_{\rm max}/V_{\rm v}) = 0.038(V_{\rm max}/V_{\rm v})^{-3.3}.
\end{equation*}
The $\vmax$ function in the Local Fields are also similarly offset (Figure~\ref{fig:vmaxdenver}), 
with the paired simulations lying 75\% higher than the isolated analogs:
\begin{equation*}
N_{0.3-1}(>V_{\rm max}) = N_0\left(\frac{V_{\rm max}}{\rm10\,km/s}\right)^{-3.1},
\end{equation*}
with $N_0 = 540$ for the paired sample and $300$ for the
isolated analogs.  Likewise, we predict similar numbers of objects with 
$\vmax > 8\,\kms$ within the 1~Mpc of each host and within the Local Volume 
around each pair as predicted in Figure~\ref{fig:MpMpc} for 
$\mpeak > 6\times10^7\msun$; these $\vmax$ functions are plotted in 
Figure~\ref{fig:vmaxmpc}.

\bibliographystyle{mn2e}
\bibliography{elvis1}

\label{lastpage}
\end{document}